\documentclass[12pt]{article}

\usepackage{latexsym}
\usepackage{amssymb}
\usepackage{amsbsy}
\usepackage{amsmath}
\usepackage{cite}
\usepackage{epsfig}
\usepackage{graphics,color}
\usepackage{a4}
\numberwithin{equation}{section}

\allowdisplaybreaks

\newcommand{\bra}{\langle}
\newcommand{\ket}{\rangle}

\newcommand{\hm}{\hspace*{-0.6cm}}

\newcommand{\be}{\begin{equation}}
\newcommand{\ee}{\end{equation}}

\newcommand{\bea}{\begin{eqnarray}}
\newcommand{\eea}{\end{eqnarray}}
\newcommand{\bean}{\begin{eqnarray*}}
\newcommand{\eean}{\end{eqnarray*}}

\newcommand{\half} {\frac{1}{2}}
\newcommand{\om} {\omega}

\newcommand{\re}{\operatorname{Re}}
\newcommand{\im}{\operatorname{Im}}

\newcommand{\rmR}{{\rm R}}
\newcommand{\rmI}{{\rm I}}

\newcommand{\lb}{\bar\lambda}

\newcommand{\nn}{\nonumber}

\begin{document}

\title{\bf\large
Localised distributions and criteria for correctness \\ in complex 
Langevin dynamics
}

\author{
 Gert Aarts$^a$\thanks{email: g.aarts@swan.ac.uk}, 
 \addtocounter{footnote}{1}
 Pietro Giudice$^{a,b}$\thanks{email:  p.giudice@uni-muenster.de}  \, and 
Erhard Seiler$^c$\thanks{email: ehs@mppmu.mpg.de} 
 \\
\mbox{} \\
{${}^a$\em\normalsize Department of Physics, College of Science, Swansea 
University}\\
{\em\normalsize Swansea, United Kingdom} \\
{${}^b$\em\normalsize Universit\"at M\"unster, Institut f\"ur Theoretische Physik} \\
{\em\normalsize  M\"unster, Germany\footnote{Present address} } \\
{${}^c$\em\normalsize Max-Planck-Institut f\"ur Physik (Werner-Heisenberg-Institut)
}\\
{\em\normalsize  M\"unchen, Germany} \\
\mbox{} \\
\mbox{} \\
}

\date{June 13, 2013}

\maketitle

\begin{abstract}
Complex Langevin dynamics can solve the sign problem appearing in numerical simulations of theories with a complex action. In order to justify the procedure, it is important to understand the properties of the real and positive distribution, which is effectively sampled during the stochastic process. In the context of a simple model, we study this distribution by solving the Fokker-Planck equation as well as by brute force and relate the results to the recently derived criteria for correctness. We demonstrate analytically that it is possible that the distribution has support in a strip in the complexified configuration space only, in which case correct results are expected.

\end{abstract}

\maketitle


\newpage
 
\tableofcontents

\section{Introduction}
 \label{sec:intro}
 
 Complex Langevin (CL) dynamics \cite{Parisi:1984cs,Klauder:1983} provides an approach to circumvent the sign problem in numerical simulations of lattice field theories with a complex Boltzmann weight, since it does not rely on importance sampling. In recent years a number of stimulating results has been obtained in the context of nonzero chemical potential, in both lower and four-dimensional field theories with a severe sign problem in the thermodynamic limit \cite{Aarts:2008rr,Aarts:2008wh,arXiv:1006.0332,Aarts:2011zn,Fromm:2012eb,Seiler:2012wz} (for two recent reviews, see e.g.\ Refs.\  \cite{Aarts:2013bla,Aarts:2013uxa}).
 However, as has been known since shortly after its inception, correct results are not guaranteed \cite{Ambjorn:1985iw,Ambjorn:1986fz,Berges:2006xc,Berges:2007nr,arXiv:1005.3468,Pawlowski:2013pje}. 
 This calls for an improved understanding, relying on the combination of analytical and numerical insight. In the recent past, the important role  played by the properties of the real and positive probability distribution in the complexified configuration space, which is effectively sampled during the Langevin process, has been clarified \cite{arXiv:0912.3360,arXiv:1101.3270}. An important conclusion was that this distribution should be  sufficiently localised in order for CL to yield valid results. 
Importantly, this insight has recently also led to promising results in nonabelian gauge theories, with the implementation of SL($N, \mathbb{C}$) gauge cooling \cite{Seiler:2012wz,Aarts:2013uxa}.
   
   The distribution in the complexified configuration space is a solution of the Fokker-Planck equation (FPE) associated with the CL process. However, in contrast to the case of real Langevin dynamics, no generic solutions of this FPE are known (see e.g.\ Ref.\ \cite{Damgaard:1987rr}). In fact, even in special cases only a few results are available \cite{Ambjorn:1985iw,Aarts:2009hn,arXiv:0912.3360,Duncan:2012tc}. 
   In Refs.\  \cite{arXiv:0912.3360,arXiv:1101.3270} this problem was addressed in a constructive manner by deriving a set of criteria for correctness, which have to be satisfied in order for CL to be reliable. These criteria reflect properties of the distribution and, importantly,  can easily be measured numerically during a CL simulation, also in the case of multi-dimensional models and field theories \cite{Aarts:2011zn}.
 
A widely used toy model to understand CL is the simple integral
\be
\label{eq:Z}
Z = \int_{-\infty}^\infty dx\, e^{-S}, 
\quad\quad\quad\quad
 S=\half\sigma x^2+\frac{1}{4}\lambda x^4,
\ee
where the parameters in the action are complex-valued.
This model has been studied shortly after CL was introduced \cite{Klauder:1985ks,Ambjorn:1985iw,Okamoto:1988ru}, but no complete solution was given. As we will see below, 
its structure, with complex $\sigma$, is relevant for the relativistic Bose gas at nonzero chemical potential \cite{Aarts:2008wh,Aarts:2009hn}. Recently, a variant of this model (with $\sigma=0$ and $\lambda$ complex) was studied by Duncan and Niedermaier \cite{Duncan:2012tc}: in particular they constructed the solution of the FPE, using an expansion in terms of Hermite functions. They considered the case of  ``complex noise'', in which both the real and imaginary parts of the complexified variables are subject to stochastic kicks. Unfortunately, it has been shown in the past that generically complex noise may not be a good idea, since it leads to broad distributions in the imaginary direction and hence incorrect results \cite{arXiv:0912.3360,arXiv:1101.3270}. This was indeed confirmed in Ref.\ \cite{Duncan:2012tc}.

In this paper we aim to combine the insights that can be distilled from the criteria for correctness discussed above with the explicit solution of the FPE, adapting the method employed in Ref.\ \cite{Duncan:2012tc} to the model (\ref{eq:Z}). The paper is organised as follows. In Sec.\ \ref{sec:cc} we discuss CL and the criteria for correctness. To keep the paper sufficiently accessible, we first briefly review how to arrive at the criteria for correctness and subsequently present numerical results, for both real and complex noise. 
In Sec.\ \ref{sec:pd} we study the probability distribution in the complexified configuration space, by solving the FPE directly as well as by a brute-force construction using the CL simulation, again for complex and real noise (the latter was not considered in Ref.\ \cite{Duncan:2012tc}). In Sec.\ \ref{sec:inter} we combine our findings concerning the distribution and the criteria for correctness, and provide a complete characterisation of the dynamics. 
Sec.\ \ref{sec:conc} contains the conclusion. Finally,  in order to see whether the structure found numerically can be understood analytically, a perturbative analysis of the FPE is given in Appendix \ref{sec:pert}.

\section{Complex Langevin dynamics and criteria for correctness}
\label{sec:cc}

We consider the partition function (\ref{eq:Z}). We take $\lambda$ real and positive, so that the integral exists, while $\sigma$ is taken complex. Analytical results are available: a direct evaluation of the integral yields
\be
 Z =  \sqrt{\frac{4\xi}{\sigma}}e^{\xi} K_{-\frac{1}{4}}(\xi),
\ee
 where $\xi=\sigma^2/(8\lambda)$ and $K_p(\xi)$ is the modified Bessel 
function of the second kind. Moments $\bra x^n\ket$ can be obtained by differentiating with respect to $\sigma$. 
Odd moments vanish.

The aim is to evaluate expectation values numerically, by solving a CL process. We start from the Langevin equation,
\be
\dot z = -\partial_z S(z) +\eta,
\ee
where the dot denotes differentiating with respect to the Langevin time $t$ and the (Gaussian) noise satisfies 
\be
\label{eq:noise}
\bra\eta(t)\eta(t')\ket=2\delta(t-t').
\ee
 After complexification,
\be
z = x+iy, \quad\quad\quad \eta=\eta_\rmR+i\eta_\rmI, \quad\quad\quad \sigma = A+iB,
\ee
 the CL equations read
\be
\dot x = K_x(x,y) + \eta_\rmR, 
\quad\quad\quad
\quad\quad\quad
\dot y = K_y(x,y) + \eta_\rmI,
\ee
with the drift terms 
\bea
K_x \equiv&&\hm -\re\partial_z S(z) =  -Ax + By - \lambda x\left(x^2-3y^2\right), \\
K_y \equiv&&\hm  -\im\partial_z S(z) = -Ay - Bx - \lambda y\left(3x^2-y^2\right).
\eea
The form of the drift terms is similar as in the Bose gas, after a reduction to a single momentum mode \cite{Aarts:2009hn}. 

The normalisation of the real and imaginary noise components follows from Eq.\ (\ref{eq:noise}) and is given by
\bea
\nn
\bra\eta_\rmR(t)\eta_\rmR(t')\ket = &&\hm 2N_\rmR\delta(t-t'), \\
\nn
\bra\eta_\rmI(t)\eta_\rmI(t')\ket = &&\hm 2N_\rmI\delta(t-t'), \\
\bra\eta_\rmR(t)\eta_\rmI(t')\ket = &&\hm 0,
\eea
with $N_\rmR-N_\rmI=1$. Here $N_\rmI\geq 0$ is a free parameter, which can be varied. In principle, expectation values should be independent of the choice of $N_\rmI$, but in practice they are not. 
Real noise amounts to $N_\rmI=0$.

Expectation values are obtained by averaging over the noise.
After this averaging, holomorphic observables evolve according to
\be
\bra O\ket_{P(t)} = \int dxdy\, P(x,y;t)O(x+iy),
\ee
where the distribution $P(x,y;t)$ satisfies the FPE
\be
\label{eq:FP}
\dot P(x,y;t) = L^TP(x,y;t),
\ee
with the FP operator
\be
\label{eq:FPop}
L^T =  \partial_x
\left( N_\rmR\partial_x-K_x\right) +
\partial_y
\left( N_\rmI\partial_y-K_y\right).
\ee
In order to justify the approach, we also consider expectation values 
with respect to a complex weight $\rho(x,t)$, 
\be
\bra O\ket_{\rho(t)} = \int dx\, \rho(x,t)O(x),
\ee
which satisfies its (complex) FPE
\be
\dot\rho(x,t) = L_0^T\rho(x,t),
\quad\quad\quad\quad
L_0^T = \partial_x\left[\partial_x+\left(\partial_x S(x)\right)\right].
\ee
This equation has a simple stationary solution, $\rho(x)\sim e^{-S(x)}$, which is the desired weight.

The task is now to show that the two expectation values $\bra O\ket_{P(t)}$ and $\bra O\ket_{\rho(t)}$ are equal, 
\be
\label{eq:eq}
 \bra O\ket_{P(t)} = \bra O\ket_{\rho(t)},
\ee
at least in the limit of large $t$, making use of the respective FPEs and the Cauchy-Riemann (CR) equations \cite{arXiv:0912.3360,arXiv:1101.3270}. Here it is essential that only holomorphic observables are considered, which evolve according to
\be
\partial_t O(z,t) = \tilde L O(z,t),
\ee
with the Langevin operator
\be
\tilde L = \left[\partial_z-\left(\partial_zS(z)\right)\right]\partial_z.
\ee
We note that for holomorphic observables, $\tilde L = L$, where $L$ is the transpose of $L^T$ introduced above. The equivalence (\ref{eq:eq}) can indeed be shown, as discussed in detail in Refs.\ \cite{arXiv:0912.3360,arXiv:1101.3270},
provided that integration by parts in $y$ is allowed,  without the presence of boundary terms at infinity. 
This construction involves the products $P(x,y;t)O(x+iy)$ for `all' observables $O(x)$, and hence it puts severe constraints on the decay of the distribution at infinity. This will indeed be shown to be crucial below.

\begin{figure}[t]
\begin{center}
\epsfig{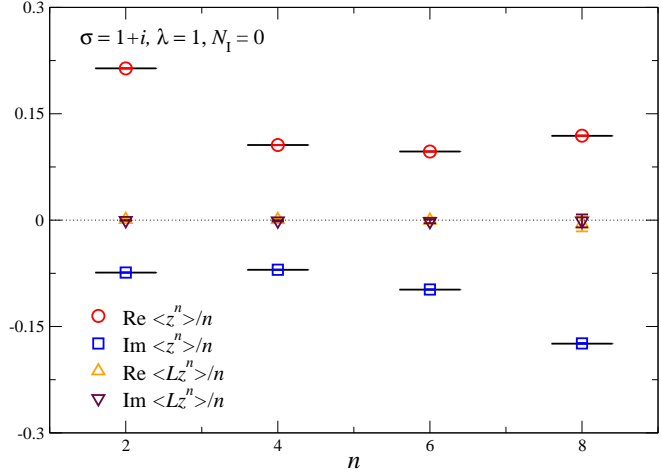}
\end{center}
 \caption{Real and imaginary parts of the expectation values $\frac{1}{n}\bra z^n\ket$ and criteria for correctness  $C_n=\frac{1}{n}\bra \tilde Lz^n\ket$ versus $n$ at $\sigma=1+i$ and $\lambda=1$ for real noise ($N_\rmI=0$).  The horizontal lines indicate the exact value. 
  }
 \label{fig:CC}
\end{figure}

From now on we consider only the equilibrium distribution $P(x,y)$, assuming that it exists, and hence drop the $t$ dependence. In the large $t$ limit, the equivalence (\ref{eq:eq}) can then be expressed in terms of the {\em criteria for correctness} \cite{arXiv:0912.3360,arXiv:1101.3270}
\be
C_O \equiv \left\bra\tilde L O(z)\right\ket=0,
\ee
which in principle need to be satisfied for a complete set of observables $O(z)$. Here the expectation value is taken with respect to the equilibrium distribution $P(x,y)$, or equivalently, a noise average. 
After separating real and imaginary parts, the criteria take the form
\bea
\re \tilde L O = &&\hm \re O''+K_x\re O' - K_y \im O', \\
\im \tilde L O = &&\hm \im O''+K_x\im O' + K_y \re O',
\eea
where the primes denote differentiation with respect to $z$.
 We consider as observables
\be
O_n(z) = \frac{1}{n}z^n,
\ee
with $n$ even (the odd powers vanish by symmetry). The associated consistency conditions,
\be
C_n \equiv  \frac{1}{n}\left\bra \tilde L z^n\right\ket = 0,
\ee
then take the explicit form
\bea
\label{eq:C2}
&& C_2 = 1-\bra\sigma z^2+\lambda z^4\ket, \\ 
&& C_4 = \bra 3 z^2-\sigma z^4-\lambda z^6\ket, \\ 
&& C_6 = \bra 5 z^4-\sigma z^6-\lambda z^8\ket, \\
&& \ldots \nn
\eea
which are of course nothing but the standard Schwinger-Dyson (SD) relations between $n$-point functions, which should be satisfied in order for the theory to be solved correctly.

\begin{figure}[t]
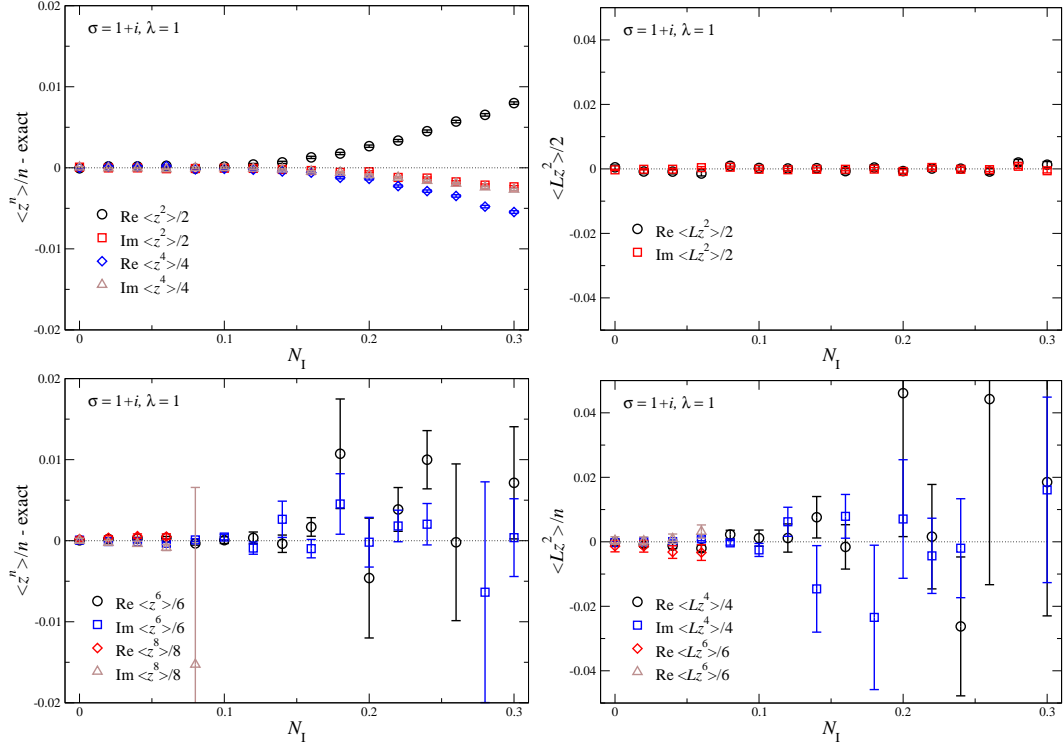

\begin{center}
\epsfig{figure=./figures/plot-obs-A1B1l1vsNI-v6.eps,width=0.48\textwidth}
\epsfig{figure=./figures/plot-CC-A1B1l1vsNI-v4.eps,width=0.48\textwidth}
\epsfig{figure=./figures/plot-obs-A1B1l1vsNI-v7.eps,width=0.48\textwidth}
\epsfig{figure=./figures/plot-CC-A1B1l1vsNI-v5.eps,width=0.48\textwidth}
\end{center}
 \caption{Observables minus the exact result (left) and criteria for correctness (right) as a function of $N_\rmI$
  at $\sigma=1+i$ and $\lambda=1$, for small $n$ (above) and larger $n$ (below).  }
 \label{fig:obs}
\end{figure}

We now turn to the numerical solution of the CL process, using the simplest lowest-order discretisation with an adaptive stepsize \cite{arXiv:0912.0617}. For the results shown here, the total combined Langevin time for each parameter set is $2\times 10^6$ Langevin time units and the maximal stepsize is $5\times 10^{-5}$. We have verified that finite stepsize corrections are negligible. We have studied various combinations of $\sigma$ and $\lambda$, keeping $\re \sigma=A>0$. Here we focus on $\sigma=1+i$ and $\lambda=1$.
 In Fig.\ \ref{fig:CC} CL results are shown for the real and imaginary parts of the observables $\frac{1}{n}\bra z^n\ket$ and for the  criteria for correctness  $C_n=\frac{1}{n}\bra \tilde Lz^n\ket$, for $n=2,4,6,8$. The figure shows the result for real noise, $N_\rmI=0$: all expectation values agree with the exact result, denoted with the horizontal lines, and the criteria for correctness are all consistent with 0, as it should be.

In Fig.\ \ref{fig:obs} we show how the observables and the criteria for correctness depend on the amount of complex noise. 
In the top figures we see that for small $N_\rmI$ the observables with $n=2,4$ appear to be consistent with the exact result, while for larger $N_\rmI$ they start to deviate. 
Perhaps surprisingly, the lowest-order criterium $C_2$ is consistent with 0 for all $N_\rmI$ shown. This implies that even though $\bra z^2\ket$ and $\bra z^4\ket$ have converged to the wrong result at larger $N_\rmI$, this occurs in such a way that the condition (\ref{eq:C2}), i.e.\ the corresponding SD equation, is still satisfied. The possibility of multiple solutions to the SD equations when solving CL has been observed earlier in Ref.\ \cite{Berges:2006xc} (see also Refs.\ \cite{Pehlevan:2007eq,Guralnik:2009pk}).

In order to detect problems, it is necessary to consider higher moments. In Fig.\ \ref{fig:obs} (below), we observe that for small $N_\rmI$ the observables (with $n\geq 6$) and the criteria (with $n\geq 4$) are only marginally consistent with the expected results, while for larger $N_\rmI$ they suffer from large fluctuations and can no longer be sensibly determined.
According to the analytical justification \cite{arXiv:0912.3360,arXiv:1101.3270}, this implies that the results from 
CL cannot be trusted in the presence of complex noise. Below we give an interpretation of this in terms of the properties of the probability distribution.
For now we tentatively conclude that, if we assume that the large fluctuations reflect the slow decay of the distribution in the imaginary direction, $P(x,y)$ should decay as $1/|y|^\alpha$, with $5 \lesssim \alpha\lesssim 7$, which will indeed be confirmed below.

\section{Probability distributions}
\label{sec:pd}

A crucial role in the justification of the method is played by the equilibrium distribution $P(x,y)$ in the complexified space. In Refs.\ 
 \cite{arXiv:0912.3360,arXiv:1101.3270} it was shown in detail that for CL to give correct results, it is necessary that the product of the distribution and a suitable basis of observables drops off fast enough in the imaginary direction. This condition can be translated into the criteria for correctness, as discussed above. Unfortunately the Fokker-Planck equation, satisfied by the distribution, cannot be solved easily, except in the case of a noninteracting model ($\lambda=0$), see Appendix \ref{sec:pert}. 

In this section we study the distribution following two approaches. Firstly, it is possible to collect histograms of the (partially integrated) distribution during the CL evolution. Note that very long runs are required, in order to sample the configuration space properly.
Here we will in particular be interested in the partially integrated distributions
\be
\label{eq:Pint}
P_x(x) =  \int_{-\infty}^\infty dy\, P(x,y),
\quad\quad\quad
P_y(y) =  \int_{-\infty}^\infty dx\, P(x,y).
\ee
We note that this approach can easily be extended to multi-dimensional integrals and field theories. We refer to this as the brute force method.

Secondly, for the zero-dimensional model we consider here, it is possible to expand the distribution in terms of a truncated set of basis functions and solve the resulting matrix problem numerically, following Duncan and Niedermaier \cite{Duncan:2012tc}. We discuss this approach in the next subsection.

\subsection{Solving the Fokker-Planck equation}

We consider the eigenvalue problem
\be
-L^T P_\kappa(x,y) = \kappa P_\kappa(x,y),
\ee
where the FP operator $L^T$ was given in Eq.\ (\ref{eq:FPop}) and takes the explicit form
\bea
L^T 
=&&\hm
 N_\rmR\partial^2_x +(Ax-By)\partial_x 
 + 
 N_\rmI\partial^2_y +(Ay+Bx)\partial_y 
 +2A
 \nn \\
  &&\hm + 
  \lambda  \left(x^3-3xy^2\right)\partial_x
+
 \lambda  \left(3x^2y-y^3\right)\partial_y
+ 6\lambda  \left(x^2-y^2\right).
\label{eq:FPop2}
\eea
We denote the eigenvalues of $-L^T$ with $\kappa$ and the 
eigenfunctions with $P_\kappa(x,y)$. If there is a unique ground state $P_0$ with 
eigenvalue $\kappa=0$, and for all other eigenvalues $\re\kappa>0$, the time-dependent distribution can be written as
\be
P(x,y;t) = P_0(x,y) + \sum_{\kappa\neq 0}  e^{-\kappa t}  P_\kappa(x,y),
\ee
and the equilibrium distribution is given by $P_0(x,y)$. In the CL simulations we observe convergence to well-defined expectation values (at least for the low moments, $n=2,4$) and hence we are certain that an equilibrium distribution exists. 

In order to solve the eigenvalue problem, we follow closely Ref.\ \cite{Duncan:2012tc}. 
The FP operator is invariant under $x\to -x, y\to -y$, which implies that eigenfunctions have 
a definite parity, $P_\kappa(x,y) = \pm P_\kappa(-x,-y)$. The ground state is expected 
to satisfy $P_0(x,y) = P_0(-x,-y)$, such that observables of the type 
$\bra (x+iy)^n\ket$, with $n$ odd, vanish. 
If $P_\kappa$ is an eigenfunction of $L^T$ with eigenvalue $\kappa$, then so is 
$P_\kappa^*$ with eigenvalue $\kappa^*$. It is expected that $P_0$ is real.

In Ref.\ \cite{Duncan:2012tc} $P(x,y)$ was doubly expanded in a basis of Hermite functions, i.e.\
\be
P(x,y) = \sum_{k=0}^{N_H-1}\sum_{l=0}^{N_H-1} c_{kl} H_k\left(\sqrt{w}x\right) H_l\left(\sqrt{w}y\right),
\ee
where $\om$ is a variational parameter appearing in the harmonic oscillator eigenfunctions, and $N_H$ indicates the number of Hermite functions included in the truncated basis. The coefficients $c_{kl}$ have to be determined.

In order to do so, we introduce creation and annihilation operators, satisfying
\be
[a,a^\dagger]=[b,b^\dagger]=1,
\ee
and write
\begin{align}
x=\frac{1}{\sqrt{2\om}}\left(a+a^\dagger\right), && p_x=-i\partial_x=i
\sqrt{\frac{\om}{2}}\left(a^\dagger-a\right),
\\
y=\frac{1}{\sqrt{2\om}}\left(b+b^\dagger\right), && p_y=-i\partial_y=i
\sqrt{\frac{\om}{2}}\left(b^\dagger-b\right).
\label{eq:ab}
\end{align}
 In terms of these, $-L^T$ reads
\bea
-L^T = &&\hm
    N_\rmR p_x^2 + N_\rmI  p_y^2- i \left( Ax-By\right) p_x  - i \left( Ay+Bx  
\right) p_y 
     -2A  
     \nn \\ &&\hm
  -6\lambda\left(x^2-y^2 \right)     +\frac{\lambda}{4\om}\left[ X(x,y)-X(y,x)
\right],
  \eea
with the quartic terms
\be
X(x,y)=   - 4i \om \left(x^3-3xy^2\right)  p_x, 
\quad\quad
X(y,x) =  - 4i \om \left(y^3-3x^2y\right) p_y.
\ee
Note that $X$  is independent of $\om$. Finally, in terms of the creation/annihilation operators, the FP operator reads
\bea
-\frac{2}{\om}L^T 
=  &&\hm
- N_\rmR \left(a^\dagger + a^2 - 2 a^\dagger a -1 \right) 
- N_\rmI \left( b^\dagger + b^2 - 2 b^\dagger b -1 \right)
\nn \\ &&\hm
+ \bar A \left( a^{\dagger 2} - a^2 + b^{\dagger 2} - b^2 +2 \right)
+ 2 \bar B  \left( b^\dagger a - a^\dagger b \right)
-4 \bar A 
\nn \\ &&\hm
- \bar\lambda  \left[ \left( a^{\dagger 2} + a^2+2 a^\dagger a \right)
- \left( b^{\dagger 2} + b^2+2 b^\dagger b \right) \right]
\nn \\ &&\hm
+ \frac{\lb}{12}  \left[ X(a,b)-X(b,a)\right],
\label{eq:fpab}
\eea
where
\be
X(a,b) =  \left(a+a^\dagger\right)^3\left(a^\dagger-a\right)
 -3 \left(a^\dagger+a\right)\left(a^\dagger-a\right) \left(b^\dagger +b \right)^2,
 \ee
and we introduced the rescaled parameters,
\be
\bar A = \frac{A}{\om}, 
\quad\quad\quad
\bar B = \frac{B}{\om}, 
\quad\quad\quad
\lb = \frac{6\lambda}{\om^2}.
\ee
In Ref.\ \cite{Duncan:2012tc}, where $A=B=0$, $\om$ was chosen to be proportional to $\sqrt{\lambda}$, and no adjustable parameters were left  on the RHS of Eq.\ (\ref{eq:fpab}). As we see below, there is a great advantage in keeping $\om$ arbitrary.

We can now compute the matrix elements with respect to the Hermite functions, using the notation
\be
|mn\ket = \frac{1}{\sqrt{m!n!}} a^{\dagger m} b^{\dagger n}|0\ket, 
\quad\quad\quad\quad
 a|0\ket=b|0\ket =0,
\ee
where
\be
H_m(\sqrt{\om}x) = \bra x|m\ket,
\quad\quad\quad\quad
H_n(\sqrt{\om}y) = \bra y|n\ket.
\ee
The matrix elements are
\bea
   -\frac{2}{\om}\bra kl | L^T |mn\ket 
 =    &&\hm
        \left[  \left(N_\rmR-\lb\right) \left(2m+1\right) 
       + \left(N_\rmI +\lb\right) \left( 2n+1\right) -2 \bar A \right]\delta_{k,m}
\delta_{l,n}
       \nn \\ &&\hm
       - \left[  \left(N_\rmR+\lb -\bar A \right)   f_{km}\delta_{k,m+2} 
        +
      \left(N_\rmR+\lb +\bar A \right)   f_{mk}\delta_{k,m-2} \right] \delta_{l,n}
\nn \\ &&\hm
    -  \left[  \left(N_\rmI -\lb -\bar A \right)   f_{ln}\delta_{l,n+2} 
        + \left(N_\rmI -\lb +\bar A \right)   f_{nl}\delta_{l,n-2}   \right] \delta_{k,m}
    \nn \\ &&\hm
+ 2\bar B \left(  \sqrt{ml }\delta_{k,m-1} \delta_{l,n+1} -   \sqrt{kn }
\delta_{k,m+1} \delta_{l,n-1}\right)  
\nn \\ &&\hm
   + \frac{\lb}{12}  \left[ X_{kl,mn}-X_{lk,nm}\right],
    \eea
 with
 \bea
 X_{kl,mn}           = &&\hm 
 \Big[   
f_{km}\delta_{k,m+4}+(2m+3-6n) f_{km} \delta_{k,m+2} +6(m-n)\delta_{k,m}
 \nn \\ &&\hm
-(2m-7-6n) f_{mk}\delta_{k,m-2}
-  f_{mk} \delta_{k,m-4}
  \Big]\delta_{l,n} 
 \nn  \\ &&\hm 
 -3\left[ 
 f_{km}\delta_{k,m+2}  -  f_{mk}\delta_{k,m-2} +\delta_{k,m}
      \right]
 \left[ 
  f_{ln}\delta_{l,n+2}  +  f_{nl}\delta_{l,n-2} 
      \right],
 \eea
and
\be
f_{km} = \sqrt{\frac{k!}{m!}}.
\ee

Following Ref.\ \cite{Duncan:2012tc}, the double 
indices $k,l$ and $m,n$ (all taking values from 0 to $N_H-1$) are converted into single ones, via
\be
 i=kN_H+l+1, \quad\quad\quad\quad     j=mN_H+n+1,  
 \ee
  and the inverse
 \begin{align}
 k_i=(i-1-\mbox{mod}(i-1,N_H))/N_H,  &&  l_i=\mbox{mod}(i-1,N_H), \\
 m_j=(j-1-\mbox{mod}(j-1,N_H))/N_H,  &&  n_j=\mbox{mod}(j-1,N_H), 
 \end{align}
with $i,j=1,\ldots , N_H^2$. The matrix elements are denoted as 
$L^T_{ij} = \bra kl | L^T |mn\ket$, and the eigenvalue problem is written as
\be
-L_{ij}^T v_j^{(\kappa)} = \kappa v_i^{(\kappa)}.
\ee
We have solved this matrix problem with a FORTRAN90 code using subroutines provided by the LAPACK library \cite{LAPACK}. Since the matrix size is $N_H^2\times N_H^2$, there is an upper limit of what is practically feasible.  For the maximal number of Hermite functions we have considered, $N_H=150$, the numerical computation takes around 36 hours on a standard work station.
Convergence can be tested by increasing $N_H$ and varying $\om$ (see the detailed discussion below). Considering the eigenvalue at  (or closest to) 0, the distribution $P_0(x,y)$ can be reconstructed from the corresponding eigenvector, as
\be
\label{eq:Prec}
P_0(x,y) = \sum_{i=1}^{N_H^2} v_i^{(0)} H_{k_i}\left(\sqrt{w}x\right) H_{l_i}\left(\sqrt{w}y\right).
\ee
Below we drop the subscript `0'.

\subsection{Complex noise}

\begin{table}[t]
\begin{center}
\begin{tabular}{|l|l|}
\hline
    $N_\rmI$ & $\om$ \\
\hline
  0      	&   3, 4, 5, 10, 40, 50, 60  \\ \hline
  0.01   	& 1.5, 2, 4, 8, 12, 16, 20 \\ \hline
 1      	& 0.5, 1, 1.5, 2, 5, 10 \\ \hline
\end{tabular}
\end{center}
\caption{Values of $N_\rmI$ and $\om$ used, with $\sigma=1+i$, $\lambda=1$, and $30\leq N_H\leq 150$.}
\label{tab:params}
\end{table}

We start with the case of complex noise. The parameters in the action are taken as $\sigma=1+i$ and $\lambda=1$, and we consider a basis with $30\leq N_H\leq 150$ Hermite functions.
The values of $\om$ we used are listed in Table \ref{tab:params}.
 In the limit of large $N_H$ the results are expected to be independent of the value of $\om$. In practice however, we find that for finite $N_H$ the parameter $\om$ plays the role of a tuning parameter: 
in particular, when $\om$ is too small, there are eigenvalues with a negative real part. This becomes more prominent as $N_\rmI$ is reduced, see below. Obviously, in this application this would mean that the FP evolution would not thermalise and display runaway behaviour. Since the CL evolution thermalises (and is obviously independent of the choice of $\om$), 
we expect the real parts of all eigenvalues to be nonnegative. When the value of $\om$ is increased,  we observe that the eigenvalues with a real negative part move into the positive half-plane and the spectrum around the origin converges. Convergence can also be seen by studying the reconstructed probability distribution $P(x,y)$, using Eq.\ (\ref{eq:Prec}).
 Interestingly, we always find an eigenvalue consistent with 0.  When $\om$ is increased even more, convergence properties worsen again. We find therefore that there is an $\om$ interval for which:
\begin{enumerate}
\item there is an eigenvalue consistent with 0;
\item the other eigenvalues are in the right half-plane;
\item the reconstructed ground state distribution is stable under variation of $N_H$ and  $\om$. 
\end{enumerate}
The $\om$ interval depends on the parameters and is pushed to larger values as $N_\rmI$ is reduced.
We have not found a special role for the $\om$ value used in Ref.\ \cite{Duncan:2012tc}, namely $\om=\sqrt{3\lambda}$ (in our conventions).

\begin{figure}[t]
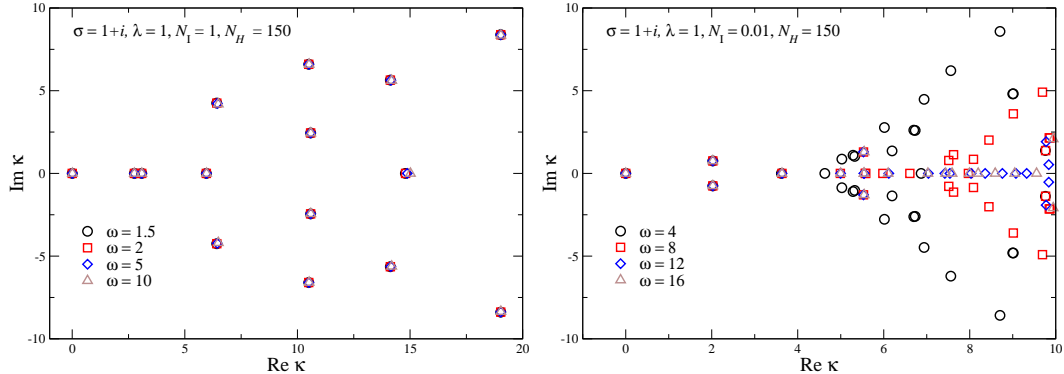

\begin{center}
\epsfig{figure=./figures/plot-kappa-A1B1l1-NI1-NH150.eps,width=0.48\textwidth}
\epsfig{figure=./figures/plot-kappa-A1B1l1-NI0.01-NH150.eps,width=0.48\textwidth}
\end{center}
 \caption{
Eigenvalues of the FP operator $-L^T$ for complex noise, with $N_\rmI=1$ (left) and 0.01 (right), magnified around the smallest eigenvalues, for various values of $\om$, at $\sigma=1+i$, $\lambda=1$, and $N_H=150$.
 }
 \label{fig:kappa-NI1}
\end{figure}

\begin{figure}[t]
\begin{center}
\epsfig{figure=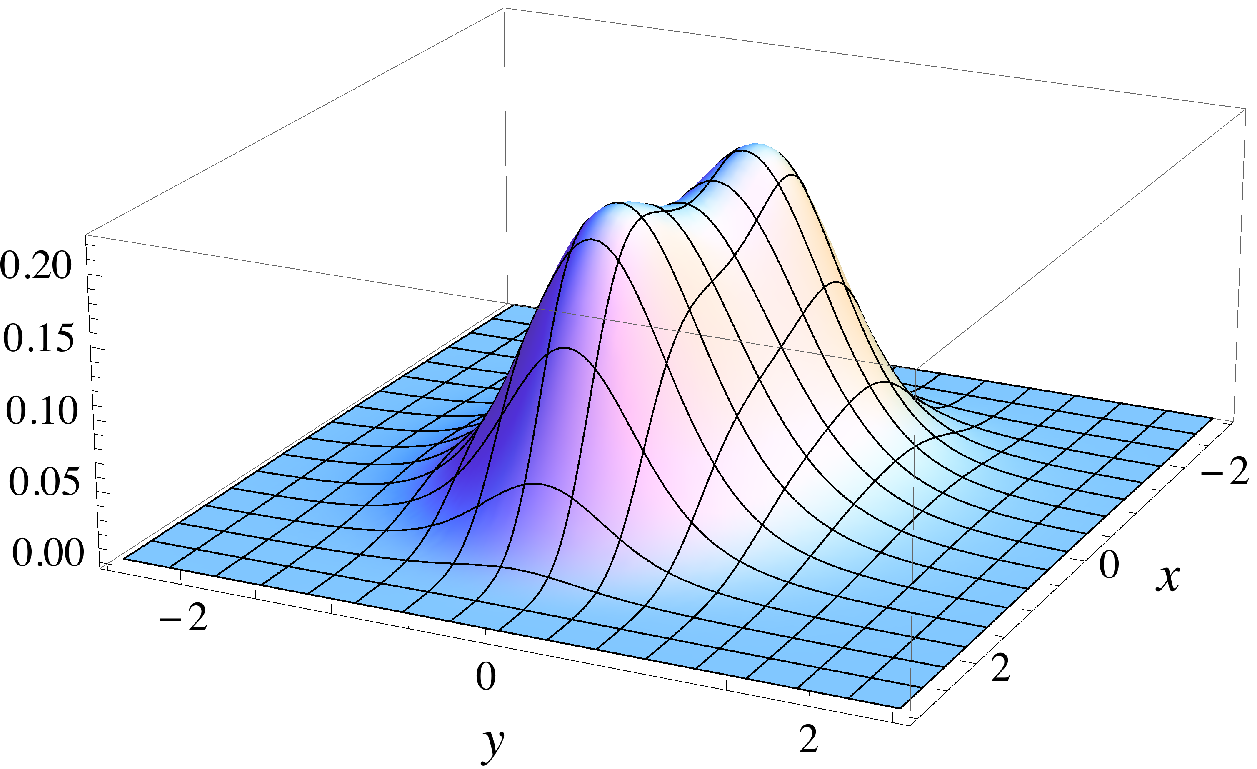, width=0.7\textwidth} \\
\vspace*{0.2cm}
\epsfig{figure=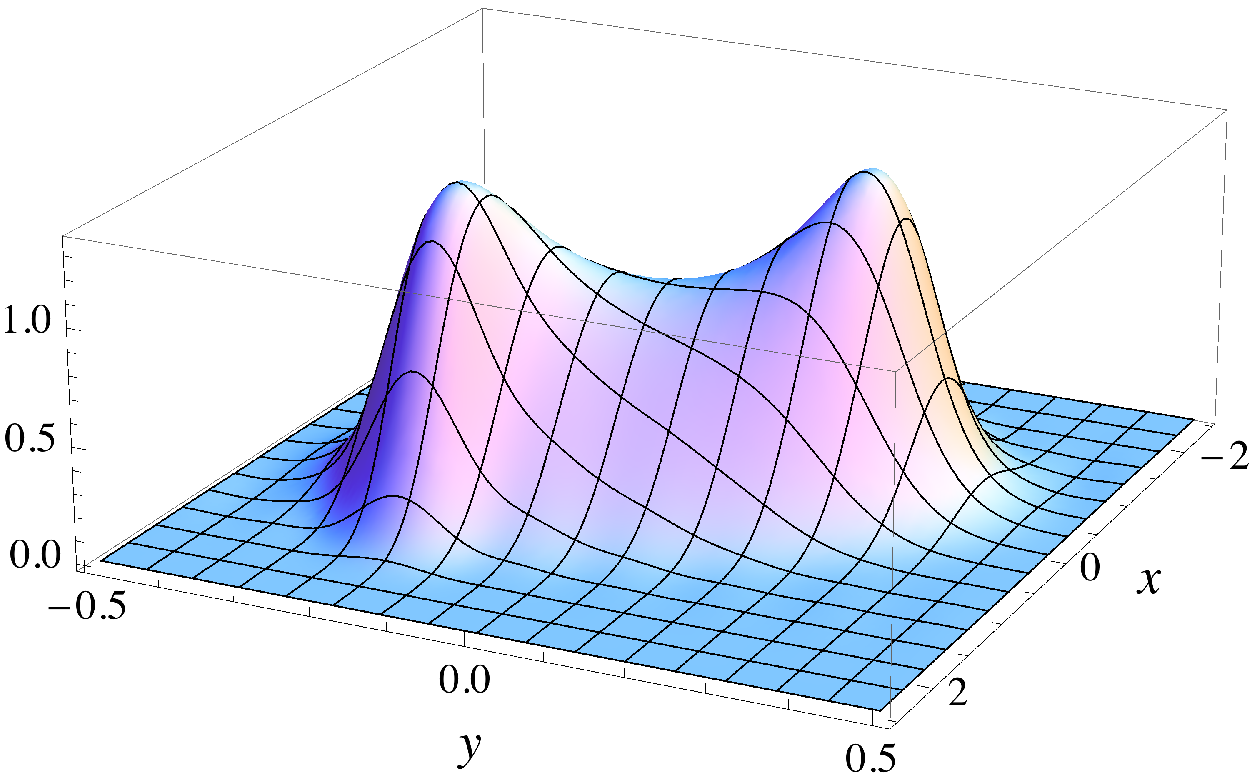, width=0.7\textwidth}
\end{center}
 \caption{
Distribution $P(x,y)$ in the $xy$-plane for complex noise, with $N_\rmI=1$ (top, with $\om=1.5$) and 0.01 (bottom, $\om=8$). Other parameters as in Fig.~\ref{fig:kappa-NI1}.}
 \label{fig:plot3d-NI1}
\end{figure}

We first consider $N_\rmI=1$, as in Ref.\ \cite{Duncan:2012tc}. 
The smallest 15 eigenvalues are shown in Fig.\ \ref{fig:kappa-NI1} (left), for several values of $\om$. For the $\om$ values shown here, all eigenvalues are in the right half-plane and the spectrum around the origin is to a good extent independent of $\om$.
The reconstructed distribution $P(x,y)$, obtained using the eigenvector corresponding to the eigenvalue at (or closest to) the origin, is shown in Fig.\ \ref{fig:plot3d-NI1} (top). We find a smooth distribution with a double peak structure, similar as in Ref.\ \cite{Duncan:2012tc}.

Next we reduce the amount of complex noise and consider $N_\rmI=0.01$. 
The spectrum is shown in Fig.\ \ref{fig:kappa-NI1} (right) and the reconstructed distribution in Fig.~\ref{fig:plot3d-NI1} (below). The findings are similar as with $N_\rmI=1$, but $\om$ has to be increased more in order to find convergence and even then the larger eigenvalues are hard to establish. The distribution has again two peaks, which are now more pronounced and rotated in the $xy$-plane. We note the symmetry $P(-x,-y)=P(x,y)$.
 Importantly, the distribution is more squeezed in the $y$ direction and the main features are contained in the interval $-0.45<y<0.45$.

\begin{figure}[t]
\begin{center}
\epsfig{figure=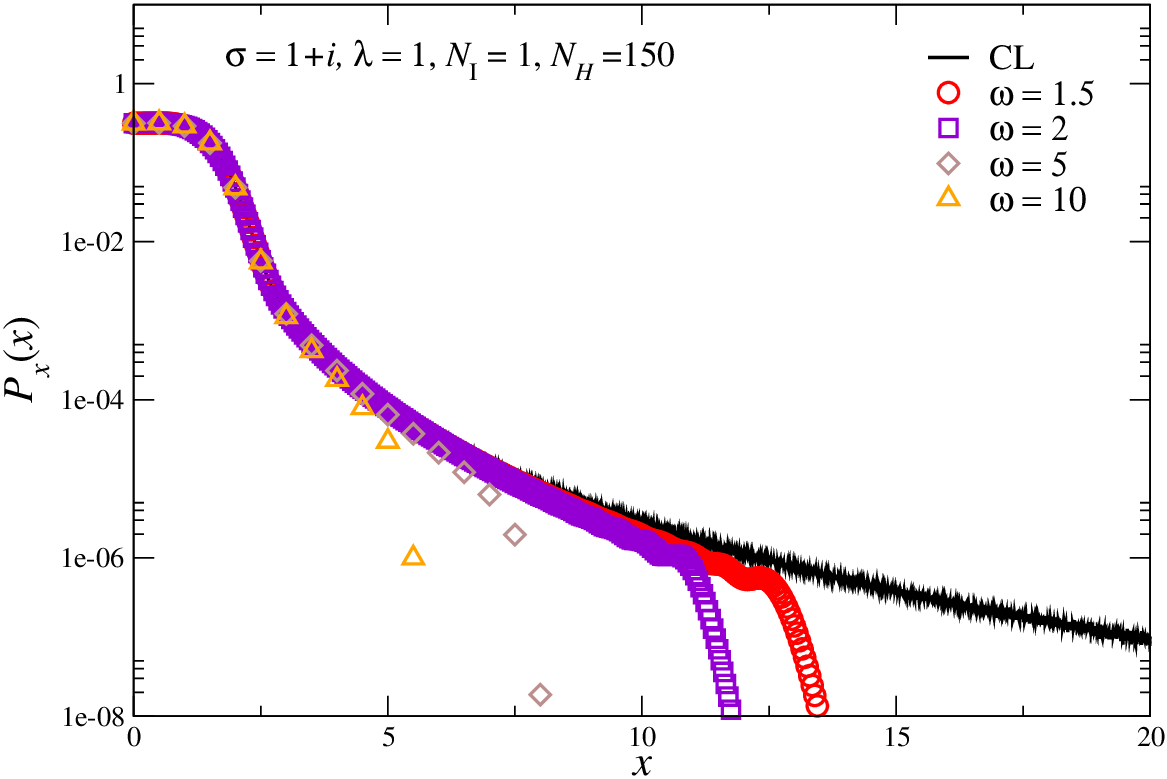,width=0.48\textwidth}
\epsfig{figure=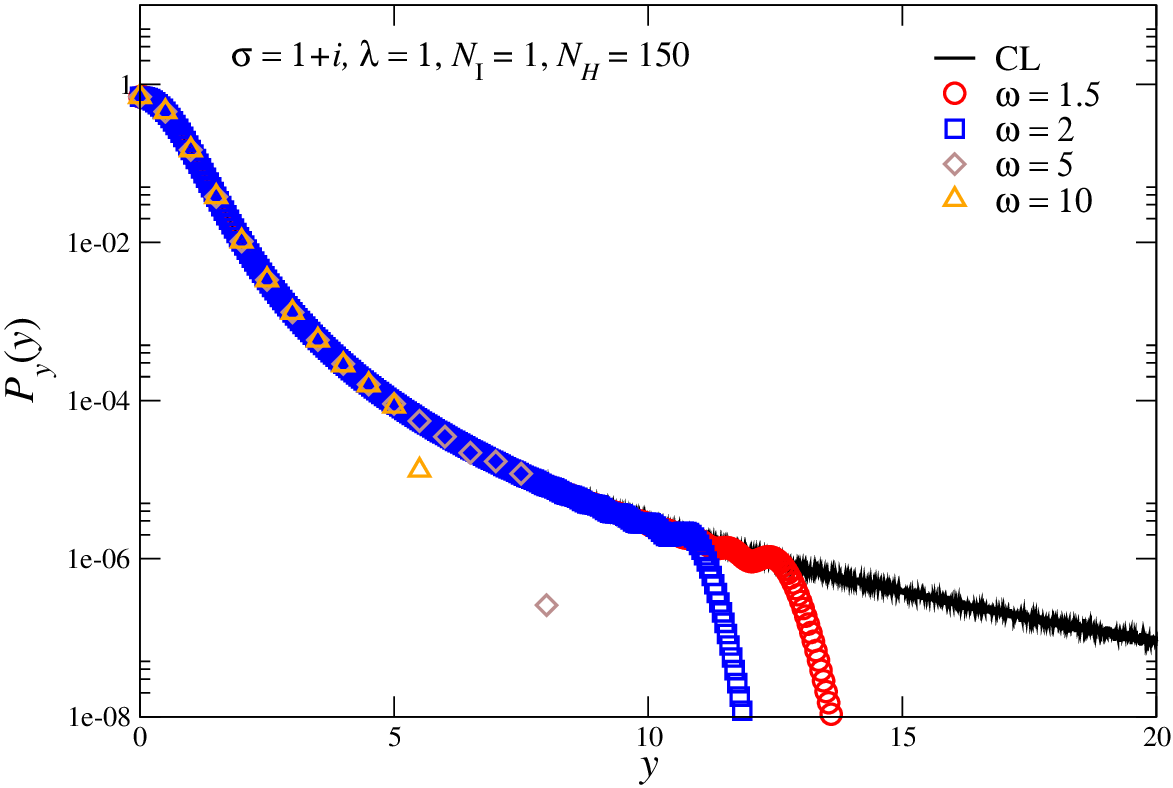 ,width=0.48\textwidth}
\end{center}
 \caption{
Partially integrated distributions $P_x(x)$  (left) and $P_y(y)$ (right) 
for different values of $\om$  with complex noise, $N_\rmI=1$. Other parameters as in Fig.\ \ref{fig:kappa-NI1}. In both cases the noisy (black) data was obtained by a CL simulation.  
 }
 \label{fig:P-NI1}
\begin{center}
\epsfig{figure=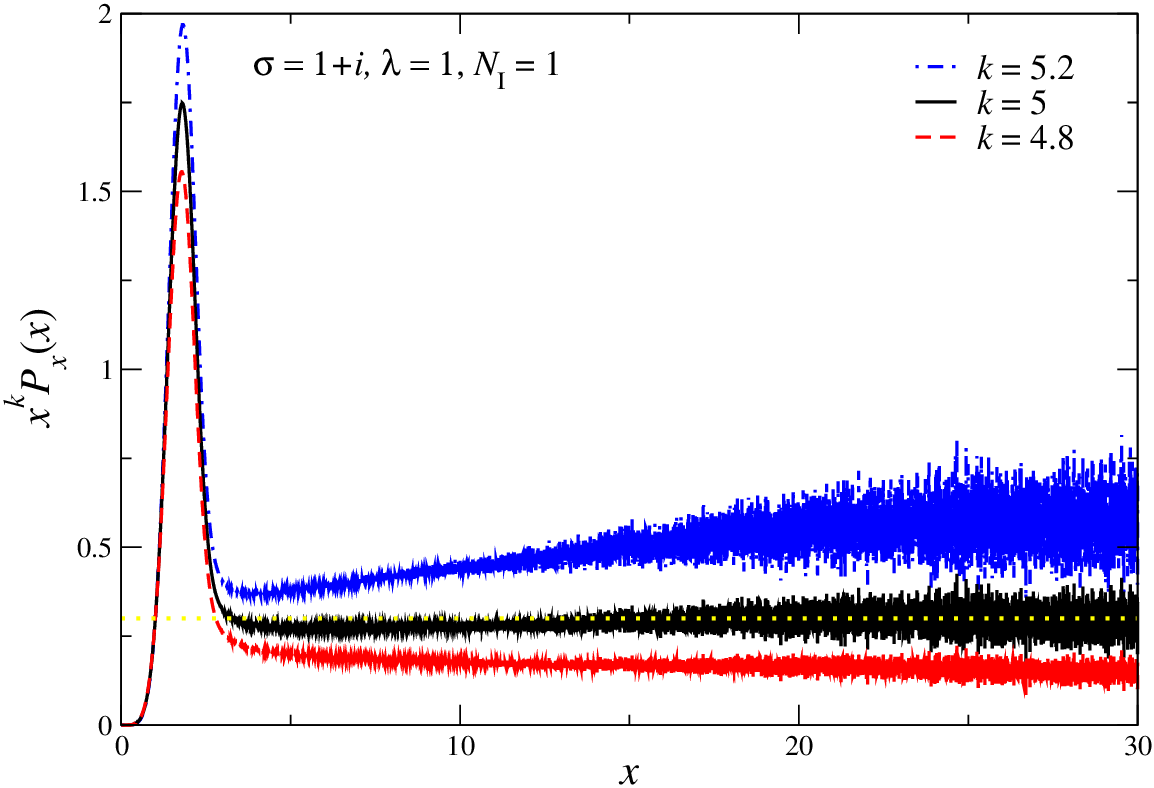,width=0.48\textwidth}
\epsfig{figure=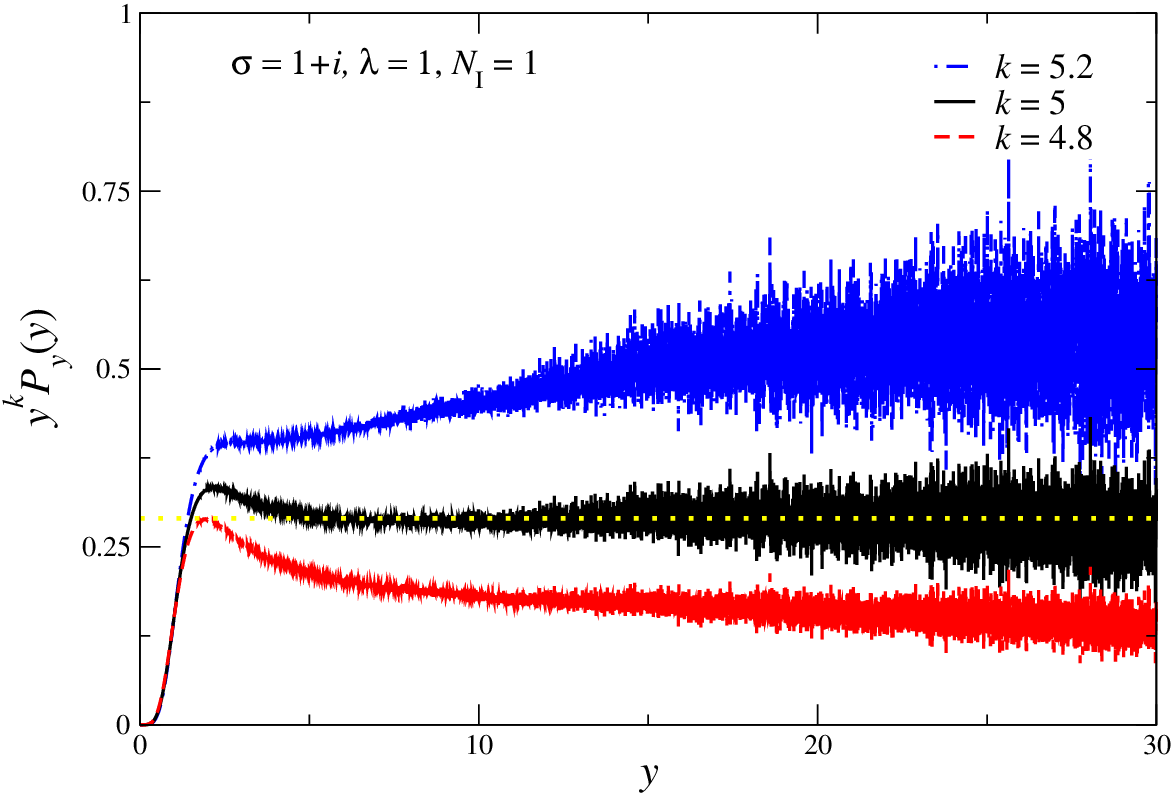 ,width=0.48\textwidth}
\end{center}
 \caption{
As above, for $x^kP_x(x)$ and $y^kP_y(y)$ with $k=4.8, 5, 5.2$, using the CL data. The dotted horizontal line is meant to guide the eye.
 }
 \label{fig:P-NI1-3}
\end{figure}

In order to clarify the relevance of these findings, we show in Fig.\ \ref{fig:P-NI1} 
the partially integrated distributions $P_x(x)$ and $P_y(y)$, see Eq.\ (\ref{eq:Pint}), on a logarithmic scale, for the case of $N_\rmI=1$. Besides presenting results for various $\om$ values, we also show  the histogram obtained during a CL simulation.
We observe an acceptable agreement between the CL results and the solution of the FPE for $\om\sim 1.5, 2$, down to a relative size of $10^{-6}$, after which the FP solution can no longer cope. We interpret this as a manifestation of the truncation. When $\om$ is taken too large, the disagreement occurs for smaller values of $x$ and $y$.

The distributions do not go to zero rapidly but decay as a power, which is clearly visible on a log-log plot. In Fig.\ \ref{fig:P-NI1-3}  we show the distributions multiplied by $x^k$ and $y^k$ respectively, for $k=4.8, 5$, and $5.2$, using the CL data. At large $|x|$ and $|y|$, we observe a power decay with power 5, i.e.
\be
P_x(x) \sim \frac{1}{|x|^5},
\quad\quad\quad\quad
P_y(y) \sim \frac{1}{|y|^5}.
\ee
This suggests that the distribution decays as
\be
P(x,y)\sim \frac{1}{(x^2+y^2)^3},
\ee
which we have verified by studying the decay of
\be
P_r(r) = \int_0^{2\pi} d\phi\, rP(r\cos\phi,r\sin\phi),
\ee
which indeed decays as $1/r^5$.
We note that this power decay is in agreement with the conclusions from the moments above: $\bra z^2\ket$ and $\bra z^4\ket$ are well defined and can be numerically determined without any problems, while the higher moments diverge, which in the CL simulation is reflected in large fluctuations.

\subsection{Real noise}

\begin{figure}[t]
\begin{center}
\epsfig{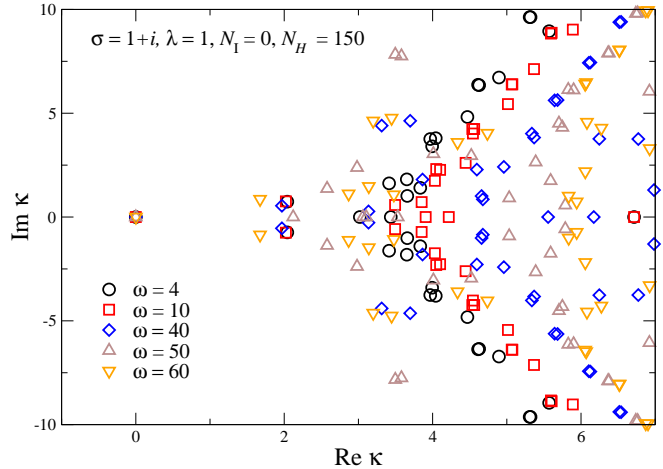}
\end{center}
 \caption{
As in Fig.\ \ref{fig:kappa-NI1}, for real noise ($N_\rmI=0$). }
 \label{fig:kappa-NI0}
\end{figure}

\begin{figure}[t]
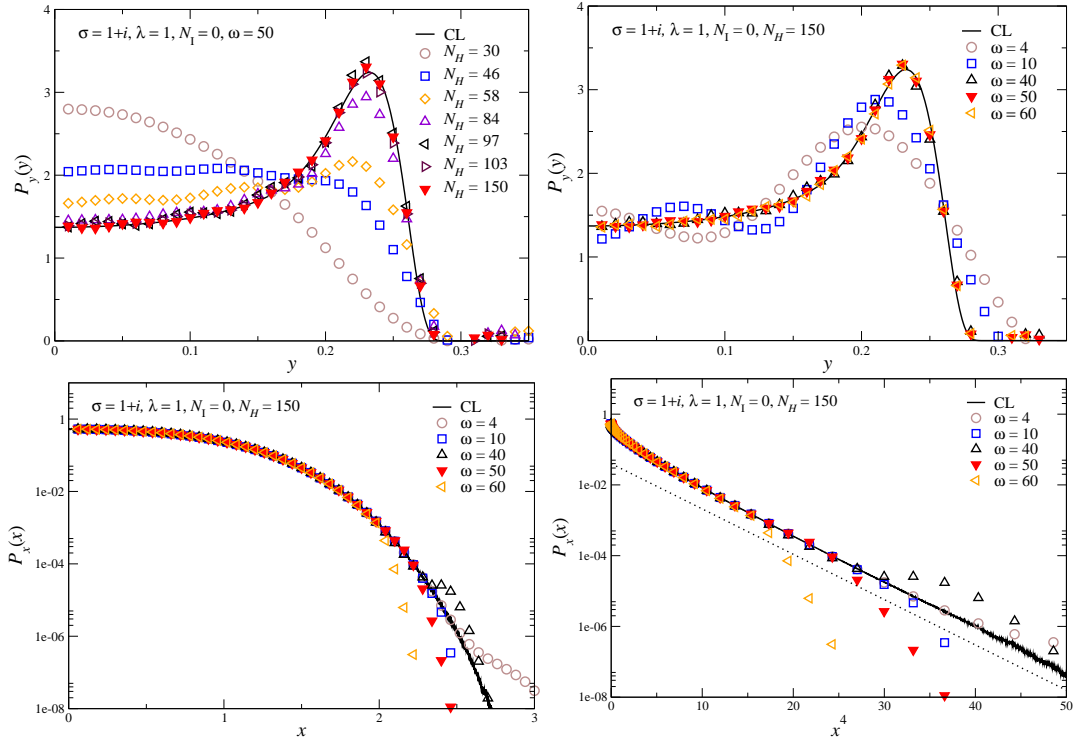

\begin{center}
\epsfig{figure=./figures/plot-Py-A1B1l1-NI0-w50-v2.eps,width=0.48\textwidth}
\epsfig{figure=./figures/plot-Py-A1B1l1-NI0-NH150-v2.eps ,width=0.48\textwidth}
\epsfig{figure=./figures/plot-Px-A1B1l1-NI0-NH150-v2.eps ,width=0.49\textwidth}
\epsfig{figure=./figures/plot-Px-A1B1l1-NI0-NH150-x4-v2.eps ,width=0.48\textwidth}
\end{center}
 \caption{
Above: Partially integrated distribution $P_y(y)$ for several values of $N_H$ and $\om=50$ (left) 
and  several values of $\om$ and $N_H=150$ (right).
Below: Partially integrated distribution $P_x(x)$ on a logarithmic scale as a function of $x$ (left) and $x^4$ (right)  for several values of $\om$ and $N_H=150$.  The dotted line on the RHS represents $P_x(x)\sim \exp(-ax^4)$ with $a=0.295$.
In both cases the black line was obtained by a CL simulation.  Other parameters as in Fig.\ \ref{fig:kappa-NI0}. 
 }
 \label{fig:P-NI0}
\end{figure}

We now turn to the case where CL appears to work well, i.e.\ with real noise ($N_\rmI=0$).
The eigenvalues are shown in Fig.~\ref{fig:kappa-NI0} for a number of $\om$ values. For $\om<4$ eigenvalues with negative real part are present (not shown in figure). We note that in all cases there is an eigenvalue at (or close to) the origin, but in general convergence is much harder to establish from a study of the eigenvalues alone. In order to have a handle on this we also analyse the partially integrated distributions $P_x$ and $P_y$ under variation of $N_H$ and $\om$, and also compare those with the histograms obtained with CL. The results are shown in Fig.\ \ref{fig:P-NI0} for $P_y(y)$ (top) and $P_x(x)$ (bottom). 
In the case of $P_y$, convergence as $N_H$ is increased is clearly visible  (top, left). 
We note that for the largest $N_H$ values the distribution  agrees with
the result obtained by direct Langevin simulation, indicated with the black line. The distribution is very well localised and appears to drop to 0 around $y=0.28$. We come back to this below.
Convergence as $\om$ is increased is demonstrated in Fig.\ \ref{fig:P-NI0} (top, right) and we observe that a large value of $\om$ is required, $\om\sim 50$. It is of course expected that the chosen value of $\om$ eventually becomes irrelevant, but for finite $N_H$ keeping $\om$ as a tuning parameter is essential.

The distribution $P_x(x)$ is shown in Fig.~\ref{fig:P-NI0} (below) as a function of $x$ (left) and $x^4$ (right), on a logarithmic scale. In contrast to the case of complex noise, we now find an exponential rather than a power decay.
Results from FPE agree with the CL histogram, independently of the value of $\omega$ in this case, but only down to a relative size of $10^{-4}$; varying $\om$ does not help in this case (increasing $N_H$ probably will).
From the CL result, we see that the distribution falls off as
\be
P_x(x) \sim e^{-ax^4},
\quad\quad\quad a\sim 0.295.
\ee
Naively this behaviour can be expected, since for large $|x|$ the original weight behaves as $\sim \exp\left(-\lambda x^4/4\right)$. We note that the prefactor is 0.295, which is slightly larger than $\lambda/4=0.25$. Interestingly this seems to be understandable from a perturbative analysis, see Appendix \ref{sec:pert}.

\begin{figure}[t]
\begin{center}
\epsfig{figure=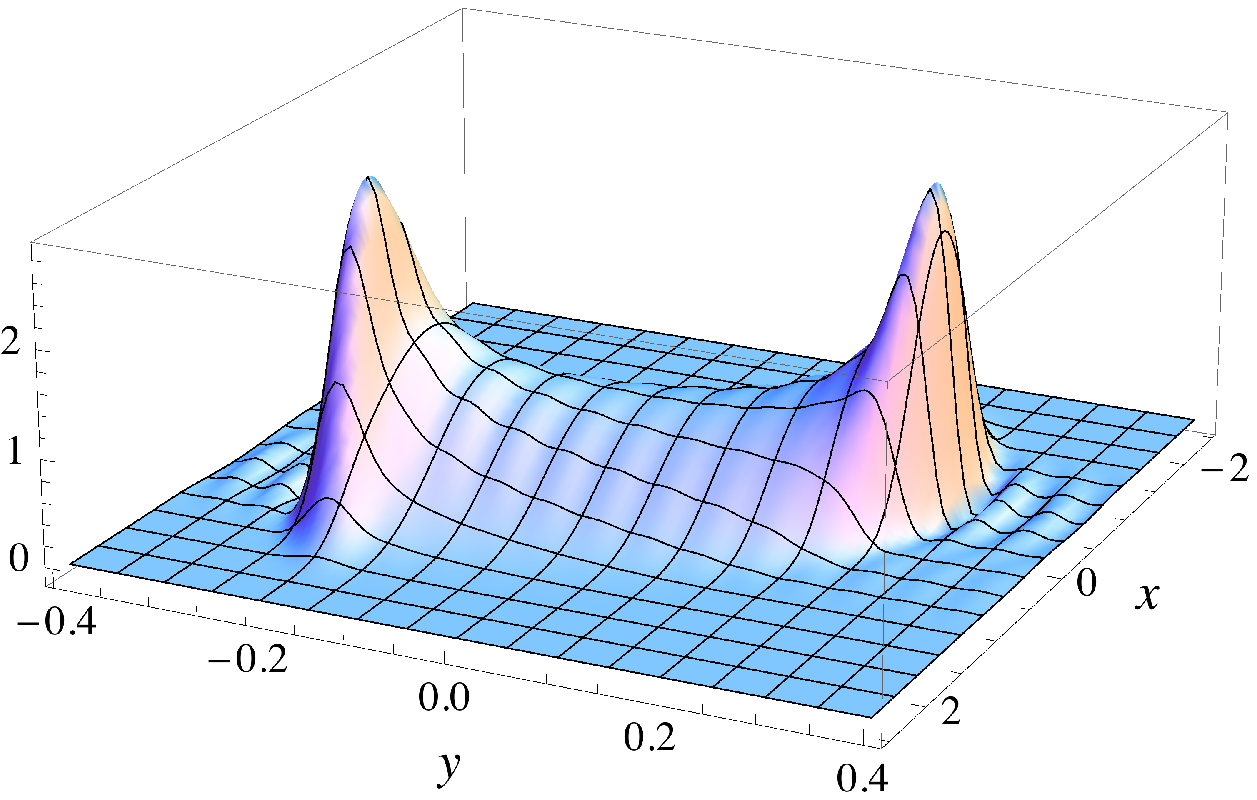 ,width=0.7\textwidth}
\end{center}
 \caption{
Distribution $P(x,y)$ in the $xy$-plane for real noise ($N_\rmI=0$) at $\sigma=1+i$ and $\lambda=1$, using $N_H=150$ and $\om=50$. 
 }
 \label{fig:plot3d-NI0}
\end{figure}

The reconstructed distribution is shown in Fig.~\ref{fig:plot3d-NI0}. This distribution has similar characteristics as at $N_\rmI=0.01$, except that the two peaks are now very pronounced and the saddle around the origin is much deeper. The peaks lie mostly in the $y$ direction and they are therefore clearly visible in $P_y(y)$. The distribution is squeezed even more than before and its main support is in the region $-0.3<y<0.3$. The ripples visible for larger $y$ values are an artefact of the truncation. In fact, in the next section we will demonstrate that the distribution is strictly 0 when $|y|>0.3029$.

We conclude that for this choice of parameters ($\sigma=1+i$ and $\lambda=1$) the decay in the case of real noise is manifestly different compared to complex noise. In the latter we found a power decay, resulting in ill-defined moments $\bra z^n\ket$ when $n>4$, while here we find exponential decay in the $x$ direction and, as we will see below,  in the $y$ direction support only inside a strip. As a result there is no problem in computing higher moments, since they are all well-defined.

\section{Interpretation}
\label{sec:inter}

From the solution of the FPE and the CL process, we conclude tentatively that for real noise the distribution is localised in the $y$ direction and has support in a strip around the origin only, with  
 $-0.3 \lesssim y\lesssim 0.3$. This conclusion can be made more precise by studying the classical flow diagram and properties of the FPE. 
This analysis can also be used to find parameter values for which CL breaks down for real noise (see Sec.\ \ref{sec:absence}).

\subsection{Classical flow}

\begin{figure}[t]
\begin{center}
\epsfig{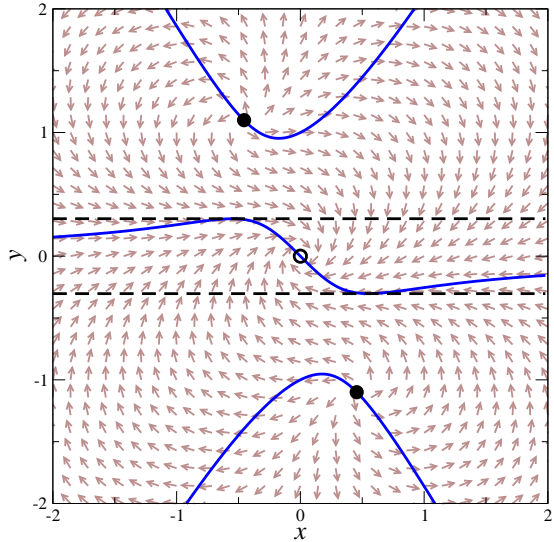}
\end{center}
 \caption{Classical flow in the $xy$-plane, for $\sigma=1+i$ and $\lambda=1$. The attractive/repulsive fixed points are indicated with the open/filled circles. The full lines indicate where $K_y(x,y)=0$. The horizontal dashed lines indicate the strip in which the CL process takes place in the case of real noise. 
 }
 \label{fig:flow}
\end{figure}

The classical flow diagram is shown in Fig.\ \ref{fig:flow}, for $\sigma=1+i$ and $\lambda=1$. We show the direction of the classical force by an arrow pointing in the direction $(K_x(x,y), K_y(x,y))$. The arrows are normalised to have the same length. The classical force is of course independent of $N_\rmI$.
There are three fixed points, where $K_x=K_y=0$: an attractive point at the origin and two repulsive fixed points, determined by $\sigma+\lambda z^2=0$, or
\be
x^2-y^2=-\frac{A}{2}, \quad\quad\quad\quad   xy=-\frac{B}{2\lambda},
\ee 
yielding $(x,y) = (\pm 0.455, \mp1.10)$ in this case.
The flow is directed towards the origin, provided that $|y|$ is not too large. This can be made more precise by studying where $K_y(x,y)$ changes sign. We find that $K_y(x,y)=0$ at
\be
y_p(x) = 2\left(\frac{B}{3\lambda}+x^2\right)^\half \cos\left(\frac{\alpha+p\pi}{3}\right),
\quad\quad\quad
p=1, 3, 5,
\ee
where
\be
\alpha = -\arctan\left(  \frac{2\lambda}{Ax} \left[\left(\frac{B}{3\lambda} +x^2\right)^3- \left( \frac{Ax}{2\lambda}\right)^2\right]^\half\right) + \pi\Theta(x),
\ee
with $\Theta(x)$ the step function. These lines are indicated in the classical flow diagram with full lines.
For the parameter values we consider here, the upper and lower curves have extrema at $x=\pm0.1749$, $y=\mp0.9530$, while the curve in the centre has its extrema at $x=\pm0.5502$, $y=\mp0.3029$. 

We now realise that along the horizontal dashed lines, which are determined by the extrema of the centre curve where $K_y=0$ ($y=\pm0.3029$ in this case), the flow is always pointing inwards, i.e.\  towards the real axis. 
In absence of a noise component in the vertical direction,
this creates a barrier for the Langevin evolution beyond which it cannot drift.
Note that the repulsive fixed points actually help to establish this. Hence, provided that the process starts within this strip,  it will never be able to leave (in the case of real noise and in the limit of zero stepsize). We have verified that if the dynamics starts out outside of the strip, it quickly finds its way into it, due to the mostly restoring properties of the classical flow.
We conclude therefore that in the case of real noise the process takes place in the strip determined by
\be
-0.3029<y<0.3029.
\ee
This is consistent with the conclusions drawn above from the histograms and the FPE solution of the distribution $P(x,y)$.
In the presence of complex noise, this conclusion no longer holds and the entire $xy$-plane can be explored.

\subsection{Strips in the complexified configuration space}

It is possible to make the argument based on classical flow presented above rigorous and show directly from the FPE that the equilibrium distribution $P(x,y)$ is strictly zero in strips in the $xy$-plane, assuming sufficient decay, i.e.
\be
 K_{x,y}(x,y)P(x,y) \to 0
\ee 
as $x$ and/or $y\to \pm\infty$.
 To achieve this, we note that the FPE takes the form of a conservation law, i.e.,
\be
\dot P(x,y;t) = \partial_xJ_x(x,y;t) +  \partial_yJ_y(x,y;t),
\ee
with
\be
J_x = \left(N_\rmR\partial_x-K_x\right) P, 
\quad\quad\quad
J_y = \left(N_\rmI\partial_y-K_y\right) P,
\ee
which allows us to consider the charge,
\be
Q(y,t) = \int_{-\infty}^\infty dx\, J_y(x,y;t).
\ee
Specialising now to the equilibrium distribution (and hence dropping the $t$ dependence), we find that $Q(y)$ is independent of $y$, provided that the product of the drift  $K_x(x,y)$ and the distribution $P(x,y)$ drops to zero at large $|x|$, since
\be
\partial_yQ(y) = \int_{-\infty}^\infty dx\, \partial_yJ_y(x,y) =  -\int_{-\infty}^\infty dx\, \partial_xJ_x(x,y) 
=  -J_x(x,y)\Big|_{x=-\infty}^\infty = 0.
\ee
We note that the required condition is always satisfied in our case, even in the case of the power decay. 
Since $Q(y)$ vanishes as $y\to\pm\infty$ (because $J_y(x,y)$ does, again relying on the sufficient decay), we find that
\be
Q(y) = \int_{-\infty}^\infty dx\, \left(N_\rmI\partial_y-K_y(x,y)\right) P(x,y) = 0.
\ee
For real noise, this yields therefore the condition
\be
Q(y) = \int_{-\infty}^\infty dx\, K_y(x,y) P(x,y) = 0,
\ee
for all $y$.
Since $P(x,y)$ is nonnegative, this condition allows us to derive the following useful property: if $K_y(x,y)$ has a definite sign as a function of $x$ for given $y$, $P(x,y)$ has to vanish for this $y$ value. As a function of $x$, $K_y(x,y)$ is a parabola with an extremum at
\be
x_0 = -\frac{B}{6\lambda y}
\ee
and a curvature of $6\lambda y$.
The value at the extremum is given by
\be
F(y) \equiv K_y(x_0,y) =  -\frac{\lambda}{y}\left[ \left( y^2-\frac{A}{2\lambda}\right)^2 - \frac{3A^2-B^2}{12\lambda^2}\right].
\ee
Consider now the case that $y$ is positive (negative). In that case, when $F(y)>0$ ($F(y)<0$), $K_y(x,y)$ is strictly positive (negative) and hence $P(x,y)$ has to vanish.
The zeroes of $F(y)$ are given by
\be
y_\pm^2 = \frac{A}{2\lambda}\left( 1\pm \sqrt{1-\frac{B^2}{3A^2}}\right),
\ee
provided that $3A^2-B^2>0$. Inspection shows that $F(y)>0$ when $y_-<y<y_+$ and $F(y)<0$ when $-y_+<y<-y_-$: hence for these $y$ values, $P(x,y)=0$. When $3A^2-B^2<0$, $F(y)$ has no zeroes and $F(y)$ and $y$ have opposite signs. In that case, $K_y(x,y)$ has no definite sign and the reasoning cannot be followed.

\begin{figure}[t]
\begin{center}
\epsfig{figure=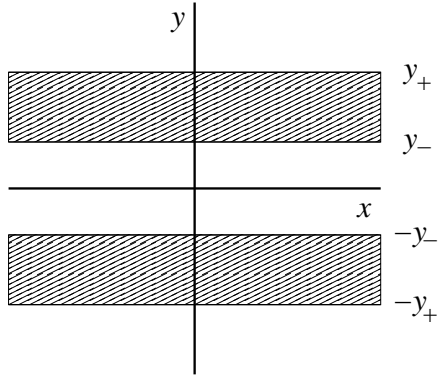, width=0.4\textwidth}
\end{center}
 \caption{
The distribution $P(x,y)$ is strictly zero in the strips bounded by $\pm y_-$ and $\pm y_+$, provided that  $3A^2>B^2$ and $N_\rmI=0$.
}
 \label{fig:strip}
\end{figure}

To summarise, we find the following:
\begin{enumerate}
\item when $3A^2>B^2$,  $P(x,y)=0$ when $y_-^2<y^2<y_+^2$, as illustrated in Fig.\ \ref{fig:strip};
\item  when $B^2>3A^2$, there are no restrictions on $P(x,y)$. 
\end{enumerate}

In the first case the distribution can in principle be nonzero in the outer region, $y^2>y_+^2$. However, once the process is in the inner strip determined by $y^2<y_-^2$, it will not be able to leave this strip, due to the nature of the drift terms. Hence there is no objection to putting the distribution to zero also when $y^2>y_+^2$. 
We conclude therefore that the equilibrium distribution has support in the strip determined by $y^2<y_-^2$ only, in agreement with the reasoning above. Note that $P(x,y)$ is therefore a nonanalytic function of $y$. 
Of course the value of $y_-$ agrees with the boundary determined in the example
in the previous section, i.e.\ with the position of the dashed lines in Fig. 10, as it should be.

For vanishing $B$, the action is real and the distribution is (for real noise) strictly localised on the real axis, $y=0$. 
For small $B$, the width of the allowed region around $y=0$ is nonzero  and set by
\be
y_-^2 \sim \frac{B^2}{12\lambda A}.
\ee
Hence increasing the amount of complexity by increasing $B$ results in a broadening of the distribution with a width $\sim 2B$. The importance of this controlled increase has been emphasised earlier in Ref.\ \cite{Aarts:2012ft}.

\subsection{Absence of strips}
\label{sec:absence}

\begin{figure}[t]
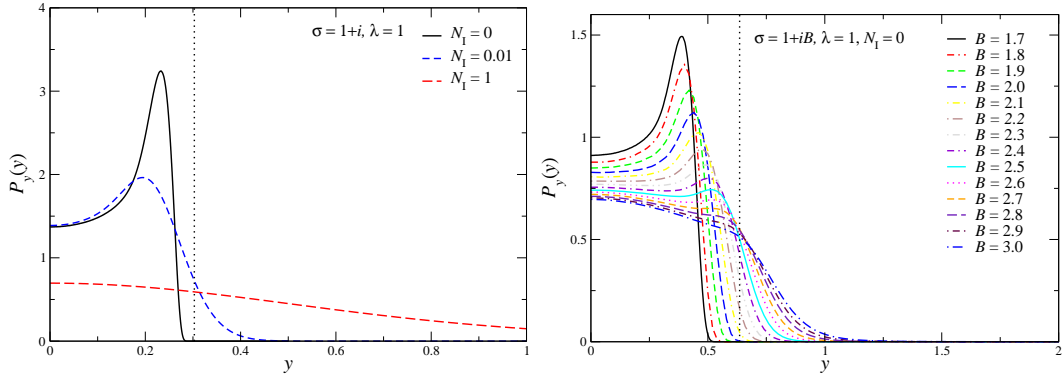

\begin{center}
\epsfig{figure=./figures/plot-Py-A1B1l1-CL.eps, width=0.48\textwidth}
\epsfig{figure=./figures/plot-Py-A1l1NI0-vsB-v2.eps, width=0.48\textwidth}
\end{center}
 \caption{
Distribution $P_y(y)$ for different values of $N_\rmI$ (left, with $B=1$) and $B$ (right, with $N_\rmI=0$)
at $\sigma=1+iB$ and $\lambda=1$, obtained with CL. On the left the vertical line at $x=0.3029$ indicates the boundary determined analytically for real noise; on the right  the vertical line indicates the boundary of the strip for $B=1.7$. For larger $B$ values, there is no longer a boundary.
 }
 \label{fig:plotPy-CL}
\end{figure}

The argument presented above breaks down in the presence of complex noise. In that case, 
the process is pushed out in the $y$ direction and the repulsive fixed points come into play.  Once the repulsive fixed point is crossed, large excursions in the $y$ direction take place and the distribution is no  longer localised.  When the amount of complex noise is small, it takes time to notice this, but eventually it will happen. There are therefore no strips for complex noise, which also follows from the formal derivation above. This is demonstrated in Fig.\  \ref{fig:plotPy-CL} (left), where $P_y(y)$ is shown for the values of $N_\rmI$ considered above. As shown above, this leads to power decay, $P_y(y)\sim 1/|y|^5$.

\begin{figure}[t]
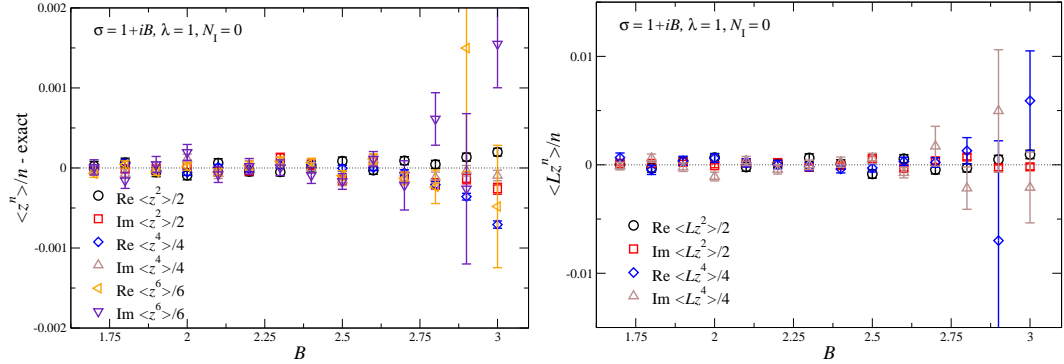

\begin{center}
\epsfig{figure=./figures/plot-obs-A1l1NI0vsB-v2.eps,width=0.48\textwidth}
\epsfig{figure=./figures/plot-CC-A1l1NI0vsB.eps,width=0.48\textwidth}
\end{center}
 \caption{Observables minus the exact result (left) and criteria for correctness (right) as a function of $B$
  at $\sigma=1+iB$, $\lambda=1$ and $N_\rmI=0$.}
 \label{fig:obsvsB}
\end{figure}

Interestingly, the derivation above demonstrates that strips are only present when $3A^2>B^2$. For larger $B$ values, one may therefore expect a breakdown of CL with real noise, similar as with complex noise. 
This is indeed what happens.
The distribution $P_y(y)$ as $B$ is increased is shown in  Fig.\  \ref{fig:plotPy-CL} (right), for real noise. Note the similarity with the figure on the left. The delocalisation  has a detrimental effect on the results of the CL process. This is demonstrated in Fig.\  \ref{fig:obsvsB}, where  the moments minus the exact result are shown on the left  and the criteria for correctness on the right. We observe that increasing $B$ has a similar effect as increasing $N_\rmI$, cf.\ Fig.\ \ref{fig:obs}.

\begin{figure}[t]
\begin{center}
\epsfig{figure=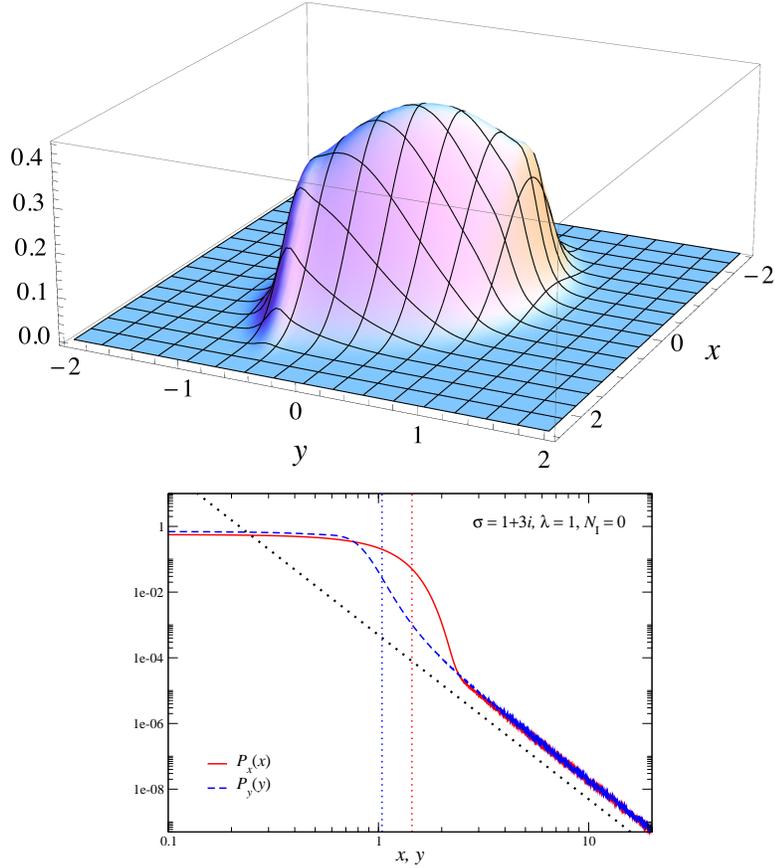, width=0.7\textwidth} \\
\vspace*{0.2cm}
\epsfig{figure=./figures/plot-PxPy-A1B3l1NI0.eps, width=0.48\textwidth}
\end{center}
 \caption{
Above: Distribution $P(x,y)$ obtained from the FPE, with $\om=6$ and $N_H=150$.
Below: Partially integrated distribution $P_x(x)$  and $P_y(y)$ on a  log-log scale, obtained from CL. The dotted line shows a power law $1/x^5$. The vertical lines indicate the $x$ and $y$ coordinate of the repulsive fixed point. In both plots, $\sigma=1+3i$, $\lambda=1$, and $N_\rmI=0$ (real noise).
}
 \label{fig:plot-PxPy}
\end{figure}

The distributions for the case that $\sigma=1+3i$ and $\lambda=1$ are shown in Fig.~\ref{fig:plot-PxPy}.
The top figure shows $P(x,y)$, obtained with the FPE. We note that the distribution still appears to be mostly contained within a strip. However, a closer look at the partially integrated distributions obtained with CL, see Fig.\ \ref{fig:plot-PxPy} (bottom), shows that again power decay is present, with the same power as before. This power decay sets in once the process has crossed the repulsive fixed points, which for this choice of parameters are located at $x=\pm1.04$ and $y=\mp 1.44$. 
The weight of the power tails is clearly small, yet it is enough to give rise to fluctuations for the higher moments when solving the CL process.
We conclude that in absence of strips a universal power law decay is present, which results in a breakdown of the formal justification \cite{arXiv:0912.3360,arXiv:1101.3270} and wrong or wildly fluctuating results in practice.

Finally we will show that it is possible to understand the universal decay directly from the FPE. 
We start from the assumption that the distribution is of the form
\be
P(x,y) = \frac{c}{(x^2+y^2)^\alpha}
\ee
at large $x$ and $y$, where we found numerically that the power $\alpha$ is consistent with 3. Substituting this Ansatz in the FPE (\ref{eq:FP}), we find, after some algebra and the removal of common factors, that
\be
\label{eq:ans}
\alpha\frac{x^2-y^2+2\alpha (N_\rmR x^2+N_\rmI y^2)}{(x^2+y^2)^2}
+ A(1-\alpha)
+ \lambda(3-\alpha)(x^2-y^2) =  0. 
\ee
At large $x$ and/or $y$ the final term dominates: requiring that this term vanishes yields indeed $\alpha=3$. 
This construction assumes that the behaviour at large distance is approximately rotationally invariant in the $xy$-plane and that there are no preferred directions, which would invalidate the Ansatz and the  power counting above. Based on our numerical evidence, this seems to be the case.
We note that the final term in Eq.\ (\ref{eq:ans}) is independent of $\sigma=A+iB$ and $N_\rmI$; hence the decay at large distance is
independent of the parameters in the action and of the amount of complex noise. We also note that $B$ has disappeared from  Eq.\ (\ref{eq:ans}): the reason is that $B$ breaks the invariance under $x\to-x$ and independently $y\to-y$, while the Ansatz is invariant under those.

The conclusion is therefore that the decay at large $x$ and $y$ is universal. Of course the presence of complex noise and/or a large value of $B^2>3A^2$ is essential in catalysing large excursions, which lead to the power decay. Notably, the power decay appears to be unavoidable unless its appearance is strictly forbidden, as in the case of the strips for real noise and $B^2<3A^2$.

\section{Conclusion}
\label{sec:conc}

In order to justify the results obtained with complex Langevin dynamics, it is necessary that the probability distribution is sufficiently localised in the complexified configuration space. Here we have studied properties of this distribution via a number of methods, in the case of a simple model. Using the insights gathered from classical flow, histograms obtained during the CL process, the criteria for correctness and the explicit solution of the FPE, a complete characterisation of the distribution can be given. 

In the case of real noise and provided that $B^2<3A^2$, where $\sigma=A+iB$, we found that the distribution is strictly localised, i.e.\ it has support in a strip in the configuration space only, with exponential decay in the real direction. In this case all moments are well-defined and, relying on the analytical proof of the method, correct results are expected. We also found that the criteria for correctness are satisfied.
In contrast, when the noise is complex or when $B^2>3A^2$, the entire configuration space is explored. Large excursions are possible due to the presence of repulsive fixed points and the decay of the distribution changes dramatically. We found strong indications that for large $|x|$ and $|y|$, the distribution decays as a power, according to 
\be
P(x,y) \sim \frac{1}{(x^2+y^2)^3}.
\ee
A consequence of this slow decay is that higher moments are no longer well-defined. As a result, these and the criteria for correctness suffer from large fluctuations during the CL process, an important signal of failure. Here it is important to emphasise that the inclusion of higher moments is essential to observe the breakdown.

In this model the FPE can be solved explicitly, via an expansion in a truncated set of basis functions. However, it is still a nontrivial problem and perhaps the best way to find the distribution is by brute force, i.e.\ during the CL simulation. This also has the benefit of being applicable to higher dimensional models. In the case of the localised distribution in the strip, the used basis set may not be the one that is best adapted to the problem and, in hindsight, once it has been demonstrated that the distribution has support in a strip only, a more suitable basis can be used. This would however limit the generality of the approach.

As an outlook, we note that in the more realistic cases of multi-dimensional models and field theories, the luxury of solving the FPE is typically not available. However, we have demonstrated that the essential insight can already be obtained from a combination of histograms of partially integrated distributions and the criteria for correctness, which gives a consistent picture of the dynamics. These tools are readily available in field theory. Finally, our conclusions are also immediately applicable to nonabelian SU($N$) gauge theories, for which gauge cooling provides a means to control the distribution in SL($N,\mathbb{C}$), a possibility not present in simpler models.

\section*{Acknowledgments}

We thank Denes Sexty and Ion-Olimpiu Stamatescu for discussion.
This work is supported by STFC and the Royal Society.

\appendix

\section{Perturbative solution of the FP equation}
\label{sec:pert}

In order to understand the numerical solution for the distribution $P(x,y)$ found above further, we discuss in this Appendix the perturbative solution of the FP equation (\ref{eq:FP}) in the stationary limit. Although it is only of limited use, it provides some insight, especially along the $x$ axis.

\subsection{Lowest-order solution}

We write the FP operator (\ref{eq:FPop}) as 
\be
L^T = L^T_0 + \lambda L^T_1,
\ee
with
\be
L^T_0 = N_\rmR\partial_x^2 +(Ax-By)\partial_x + N_\rmI \partial_y^2 +(Ay+Bx) \partial_y +2A,
\ee
and 
\be
 L_1^T  = \left(x^3-3xy^2\right) \partial_x + \left(3x^2y-y^3\right) \partial_y + 6\left(x^2-y^2\right).
\ee
The (normalisable) solution of the lowest-order equation, 
\be
L^T_0P^{(0)} = 0,
\ee
 is given by
\be
\label{eq:P0}
P^{(0)}(x,y) = N_0\exp\left[ -\alpha x^2-\beta y^2-2\gamma xy\right],
\ee
with
\bea
 \alpha = &&\hm \frac{A}{D}\left[(N_\rmR+N_\rmI)(A^2+B^2)-A^2\right], 
\\
 \beta = &&\hm \frac{A}{D}\left[(N_\rmR+N_\rmI)(A^2+B^2)+A^2\right], \\
 \gamma = &&\hm \frac{A^2B}{D}, 
 \eea
 where
 \be
 D = (N_\rmR+N_\rmI)^2(A^2+B^2)-A^2,
\ee
and $N_0$ is the normalisation constant,
\be
\frac{1}{N_0} = \int dxdy\, e^{-\alpha x^2-\beta y^2-2\gamma xy} = \frac{\pi}{\sqrt{\alpha\beta-\gamma^2}}.
\ee
This solution is similar to the one found in the relativistic Bose gas at nonzero chemical potential \cite{Aarts:2009hn}.
It is easy to see that it is the correct solution at leading order, by 
computing (recall that $N_\rmR-N_\rmI=1$ and $\sigma=A+iB$)
\be
\left\bra (x+iy)^2 \right\ket _P = \int dxdy\, P^{(0)}(x,y)(x+iy)^2 = \frac{1}{\sigma}.
\ee
More generally, one may equate the two expectation values
\bea
\bra O(x)\ket_\rho = &&\hm \int dx\, \rho(x)O(x),
\\
\bra O(x+iy)\ket_P = &&\hm  \int dxdy\, P(x,y)O(x+iy),
\eea
which, assuming that it is possible to shift $x\to x-iy$, yields the relation  \cite{Nakazato:1986dq,Namiki:1992wf}
\be
\rho(x) = \int dy\, P(x-iy,y),
\ee
where the LHS should be independent of $N_{\rmR,\rmI}$.
Evaluating the $y$ integral yields in this case
\be
\rho^{(0)}(x) = N_0' \, e^{-S(x)}, \quad\quad\quad S(x) = \half\sigma x^2,
\ee
with
\be
N_0' = \sqrt{\frac{\sigma}{2\pi}},
\ee
which is indeed the expected answer.

\subsection{First-order correction}

To compute higher-order corrections, we expand
\be
P(x,y) = \sum_{k=0}^\infty \lambda^k P^{(k)}(x,y).
\ee
Higher-order corrections are determined by the inhomogeneous partial differential equation,
\be
L_0^TP^{(k)} + L_1^TP^{(k-1)}=0.
\ee
The homogeneous equation is solved by $P^{(0)}$. To find the particular solution, we factor out the leading order solution, 
\be
 P^{(k)}=P^{(0)} p^{(k)}, 
 \ee
(with $p^{(0)}=1$), and write
\be
L_0^TP^{(k)} = P^{(0)} L_0^{\prime\, T} p^{(k)},
\quad\quad\quad
L_1^TP^{(k)} = P^{(0)} L_1^{\prime\, T} p^{(k)},
\ee 
with 
\bea
  L_0^{\prime\, T} = &&\hm 
 N_\rmR \left[ \partial_x - 4(\alpha x + \gamma y) \right] \partial_x  + (Ax-By) \partial_x 
 \nn \\ &&\hm 
 +  N_\rmI  \left[ \partial_y - 4(\beta y+\gamma x)\right] \partial_y + (Ay+Bx) \partial_y,
\\ 
  L_1^{\prime\, T} = &&\hm 
  \left(x^3-3xy^2\right)\left(-2\alpha x-2\gamma y+\partial_x\right)
    \nn \\ &&\hm
 + \left(3yx^2-y^3\right)\left(-2\beta y-2\gamma x+\partial_y\right)
 + 6\left(x^2-y^2\right).
\eea
Higher-order corrections are then determined by
\be
  L_0^{\prime\, T} p^{(k)} = - L_1^{\prime\, T}p^{(k-1)}.
\ee
For the first-order correction, this yields
\be
\label{eq:P1tilde}
 L_0^{\prime\, T} p^{(1)} = 
2\alpha x^4 +8\gamma x^3y -6(\alpha-\beta) x^2y^2 -8\gamma xy^3
 -2\beta y^4 -6\left(x^2-y^2\right).
\ee
The RHS of Eq.\ (\ref{eq:P1tilde}) is
a fourth order polynomial with only even powers. 
As a particular solution we may therefore attempt a polynomial of fourth 
degree, with only even terms appearing and containing 8 unknown 
coefficients,
\be
\label{eq:P1}
 p^{(1)}(x,y) =  c_{40} x^4 + c_{31} x^3y + c_{22} x^2y^2 + c_{13}  xy^3 + c_{04} y^4
 +  c_{20} x^2 + c_{11} xy + c_{02} y^2.
\ee
Inserting this Ansatz in Eq.\ (\ref{eq:P1tilde}) yields a set of linear equations for the coefficients which can be solved.  
Since the expressions become rather unwieldy, we give here the results for real noise only, since this is the case of interest.

 \begin{figure}[t]
\begin{center}
\epsfig{figure=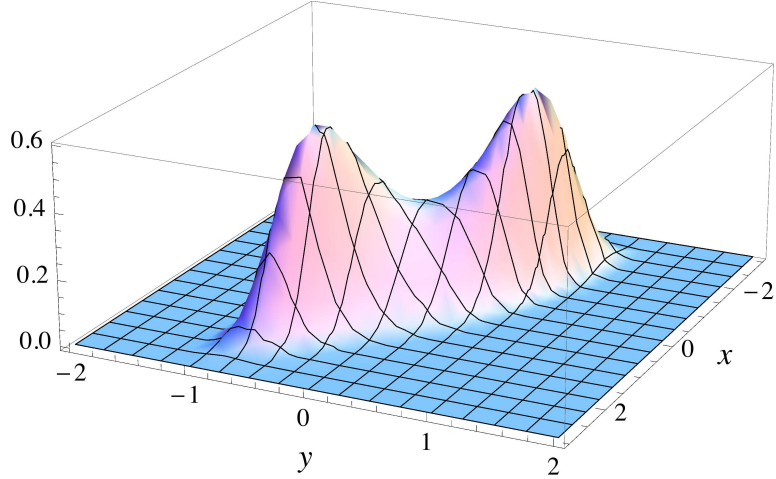 ,width=0.7\textwidth}
\end{center}
 \caption{
Distribution $P(x,y)$ in the $xy$-plane, at first nontrivial order in a perturbative expansion,
 at $\sigma=1+i$ and $\lambda=1$, for real noise. 
 }
 \label{fig:plot3d-NI1-pert}
\end{figure}

For real noise ($N_\rmR=1, N_\rmI=0$), the parameters in the lowest-order solution (\ref{eq:P0}) simplify, and 
\be
 \alpha = A, 
 \quad\quad\quad
 \beta =  A\left(1+\frac{2A^2}{B^2}\right), 
  \quad\quad\quad
 \gamma = \frac{A^2}{B}.
\ee
The coefficients of the first-order correction (\ref{eq:P1}) are given by
\begin{align}
& c_{20} = 0, 
&& c_{02} = \frac{12A(2A^2-B^2)}{B^2(4A^2+B^2)},
\\
& c_{11} = -\frac{6(A^2+B^2)}{B(4A^2+B^2)},
&& c_{22} = -\frac{9A^2(4A^2-B^2)}{B^2(4A^2+B^2)}, 
\\
& c_{40} = -\frac{3A^2}{2(4A^2+B^2)},
&& c_{04} = -\frac{A^2(36A^2-5B^2)}{2B^4}, \\
& c_{31} = -\frac{2A(5A^2-B^2)}{B(4A^2+B^2)}, 
&& c_{13} = -\frac{2A(36A^4-7A^2B^2-B^4)}{B^3(4A^2+B^2)}, 
\end{align}
Hence, to first order, the (normalised) distribution is given by
\be
P(x,y) = N_1 P^{(0)}(x,y)\left[  1 +\lambda p^{(1)}(x,y) \right],
\ee
with
\be 
\frac{1}{N_1} = 1-\frac{7A^4+3A^2B^2+2B^4)}{2(A^2+B^2)^2(4A^2+B^2)}\lambda. 
\ee
This distribution satisfies the FP equation to order ${\cal O}(\lambda)$. It can be checked that it yields the correct moments to this order, e.g.
\be
\left\bra(x+iy)^2\right\ket_P = \int dxdy\, P(x,y)(x+iy)^2 = \frac{1}{\sigma} -\frac{3\lambda}{\sigma^3} +{\cal O}(\lambda^2).
\ee
 One may also verify that evaluating
\be
\rho(x) = \int dy\, P(x-iy,y)
\ee
yields in this case
\be
\rho(x) = N_1' \, e^{-\half\sigma x^2}\left( 1-\frac{\lambda}{4}x^4\right) + {\cal O}(\lambda^2),
\ee
with
\be
N' = \sqrt{\frac{\sigma}{2\pi}}\left(1-\frac{3\lambda}{4\sigma^2}\right),
\ee
as it should be.

It is clear that the perturbative distribution is not positive definite and strictly speaking only applies when the perturbative correction $\lambda p^{(1)}(x,y)$  is small with respect to 1, i.e.\ around the origin. However, it can be made positive definite by a simple exponentiation, 
\be
 P(x,y) = P^{(0)}(x,y) \exp\left[\lambda p^{(1)}(x,y)\right],
 \ee
 which has the same leading order $\lambda$ dependence.  This distribution is normalisable since the coefficients of the quartic terms are all negative. An example is shown in Fig.\ \ref{fig:plot3d-NI1-pert}. We observe a double peak structure, as in the main text.
 
 \begin{figure}[t]
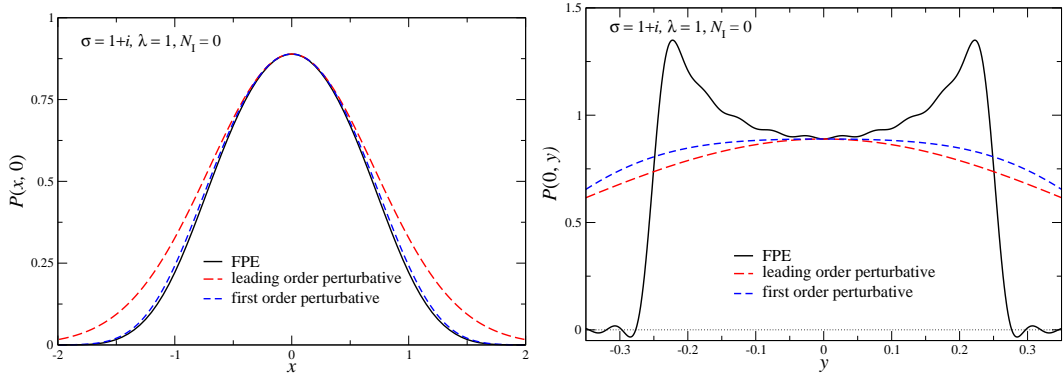

\begin{center}
\epsfig{figure=./figures/plot-Pxy-y0-vsx.eps, width=0.48\textwidth}
\epsfig{figure=./figures/plot-Pxy-x0-vsy.eps, width=0.48\textwidth}
\end{center}
 \caption{
Comparison between the perturbative distribution and the solution of the FPE, for $P(x,0)$ (left) and $P(0,y)$ (right), at $\sigma=1+i$ and $\lambda=1$, for real noise. For the solution of the FPE, $\om=50$ and $N_H=150$.
 }
 \label{fig:plot3d-NI1-pertcomp}
\end{figure}

At large $y$ values, the exponentiated construction cannot be correct, since it decays exponentially rather than be 0 outside the strip found above. In the $x$ direction, however, the perturbative solution gives a surprisingly good description of the decay. Taking $y=0$, we find
\be
P(x,0) \sim \exp(-Ax^2 + c_{40} \lambda x^4),
\ee
where, for $A=B=1$, $c_{40} = -3/10$. This result is compared with the solution of the FP equation in Fig.\ \ref{fig:plot3d-NI1-pertcomp} (left), and is seen to agree better than expected. We note that the prefactor 0.3 is also close to what was observed for the integrated distribution $P_x(x)$. 
In  Fig.\ \ref{fig:plot3d-NI1-pertcomp} (right), we also show a comparison with the perturbative expression
\be
P(0,y) \sim \exp[ -(\beta -   c_{02}\lambda) y^2 + c_{04}\lambda y^4],
\ee
where $\beta=3$, $c_{02} =12/5$ and $c_{04} = -31/2$ (again for $A=B=1$). 
Even though $c_{02}$ is positive, it  is not large enough to change the curvature. Note that the oscillations visible in the solution of the FPE are due to the finite number of basis functions ($N_H=150$).

\end{document}